\documentclass[%
  prl,
  reprint,
  amsmath,
  amssymb,
  aps
]{revtex4-2}

\usepackage{bm}
\usepackage{graphicx}
\usepackage{hyperref}

\newcommand{\vvec}[1]{\vec{#1}}

\begin{document}

% --------------------------------------------
% TITLE AND AUTHORS
\title{Score-Based Modeling of Effective Langevin Dynamics}

\author{Ludovico Theo Giorgini}
\email{ludogio@mit.edu}
\affiliation{Department of Mathematics, Massachusetts Institute of Technology, Cambridge, MA 02139, USA}

\date{\today}

% --------------------------------------------
% ABSTRACT
% --------------------------------------------
\begin{abstract}
We introduce a constructive framework to learn effective Langevin equations from stationary time series. Unlike conventional approaches that require iterative calibration to match target statistics, our construction guarantees the observed steady-state density \emph{by design} and enforces short-lag coordinate-correlation constraints directly from data---so that the surrogate satisfies the targeted invariant measure and short-lag coordinate-correlation constraints from the outset, without trial-and-error tuning. The drift is parameterized in terms of the score function---the gradient of the logarithm of the steady-state distribution---and a constant mobility matrix whose symmetric part controls dissipation and diffusion and whose antisymmetric part encodes mean nonequilibrium circulation. The score is learned from samples using denoising score matching, while the mobility coefficients are inferred from short-lag correlation identities estimated via a clustering-based finite-volume discretization on a data-adaptive state-space partition. We validate the approach on low-dimensional stochastic benchmarks and on partially observed Kuramoto--Sivashinsky dynamics, where the resulting Markovian surrogate preserves the marginal invariant measure and captures the temporal correlations of the resolved modes. The resulting Langevin models define explicit reduced generators that enable efficient sampling and generation of synthetic trajectories with the imposed statistical and dynamical constraints, without direct simulation of the underlying full dynamics.
\end{abstract}

\maketitle

% --------------------------------------------
% MAIN TEXT
% --------------------------------------------

\textit{Introduction.---}Reduced stochastic descriptions such as Langevin equations are indispensable across physics, from molecular and soft-matter systems to climate dynamics and turbulence, whenever one seeks a faithful model of a few resolved degrees of freedom while the remaining variables act as an effective bath~\cite{Risken1996,Zwanzig2001,PavliotisStuart2008,Hasselmann1976}.
Given a stationary time series $\{\vvec x(t)\}$, the inverse problem is to construct a Markovian stochastic differential equation (SDE) that reproduces relevant statistical and dynamical features of the data, such as the invariant measure and temporal correlations. These targets often suffice for key statistical predictions: matching the invariant measure ensures correct steady-state sampling of the resolved variables in stochastic dispersion models, while matching time correlations fixes effective transport rates and Green–Kubo transport coefficients, enabling surrogate-based estimation without resolving the full microscopic dynamics~\cite{Thomson1987JFM,Green1954JCP}.

Constructing such reduced models poses a fundamental challenge. Even when the underlying microscopic dynamics are Markovian, projecting out unresolved degrees of freedom generically produces a generalized Langevin equation with memory kernels and colored noise~\cite{Zwanzig1961,Mori1965}. Markovian closures can nonetheless be effective, yet they generally require careful calibration---often involving nontrivial parameter tuning---to reproduce the observed statistics across time scales, and without such tuning may fail to preserve the stationary distribution or accurately capture temporal correlations.

A broad literature addresses aspects of this inverse problem.
Classical Kramers--Moyal estimators reconstruct drift and diffusion from conditional increment moments, with finite-time corrections for coarse sampling~\cite{Siegert1998,Friedrich2000,RagwitzKantz2001}.
More recently, sparse-regression and neural approaches infer SDE models from discretely sampled data while enforcing stochastic consistency~\cite{Boninsegna2018,Callaham2021,Jacobs2023hypersindy,zhu2024dyngma,dietrich2023learning}.
Related reduced descriptions arise in molecular coarse graining (force matching/relative entropy) and transfer-operator/Markov-state approaches~\cite{IzvekovVoth2005,Shell2008,HusicNoe2018}.
In geophysical fluid dynamics, stochastic closures are often tuned to match observed covariances and lag correlations, e.g.\ via linear inverse modeling and empirical model reduction~\cite{PenlandSardeshmukh1995,Kondrashov2006,Penland89,MTV1,MTV2,KONDRASHOV201533,STROUNINE2010145,LucariniChekroun2023,KRAVTSOV,nan2019,keyes2023stochastic,giorgini2022non,falasca2025probing}.

Despite these advances, a key practical bottleneck persists in high dimensions. On one hand, methods with restrictive parametric assumptions---such as linear inverse models, which impose Gaussian steady states and linear dynamics---sacrifice fidelity: they approximate, rather than faithfully reproduce, the target statistics. On the other hand, flexible approaches that can in principle capture non-Gaussian distributions and nonlinear dynamics typically require an iterative calibration loop: once the model is specified---whether through a sparse dictionary, a neural parameterization, or a reduced closure---its parameters are commonly optimized by repeatedly evaluating model-implied transition statistics and, when long-time properties are targeted, by forward sampling until agreement with the target invariant measure and time correlations is achieved. This calibration-by-simulation becomes prohibitive when model evaluations are expensive and the state dimension is large. What is still missing is a construction in which the steady state is guaranteed \emph{a priori} and short-lag coordinate-correlation constraints are enforced directly from data, so that the resulting Langevin surrogate is correct from the outset rather than after extensive trial-and-error tuning.

In this work we introduce a constructive framework that eliminates this calibration bottleneck. Our Langevin model is assembled so that the steady-state density is satisfied \emph{by construction}, and the mobility parameters are fixed directly from data by enforcing short-lag correlation identities through a coarse generator estimate---thereby guaranteeing, from the outset, the target invariant measure together with the prescribed short-lag coordinate-correlation constraints, without repeated forward simulations. The key idea is to parameterize the drift using the steady-state density $p_{\mathrm{ss}}(\vvec x)$ through its \emph{score}, $\vvec{\nabla}\ln p_{\mathrm{ss}}(\vvec x)$, together with a constant mobility matrix whose symmetric and antisymmetric parts encode, respectively, dissipation/diffusion and mean nonequilibrium circulation.

Our construction assumes that the resolved process admits a statistically stationary distribution $p_{\mathrm{ss}}(\vvec{x})$; cyclostationary forcing can be handled by augmenting the state with the forcing phase, whereas slow nonstationary trends should be removed before applying the method. We further assume ergodicity and sufficiently fast decay of temporal correlations, so that ensemble expectations and the lagged statistics entering the inverse problem can be estimated reliably from a single sufficiently long trajectory. We assume the resolved variables to evolve on slower time scales than the unresolved degrees of freedom, which enter only through an effective stochastic influence, and are fully observed; for partial observations, the same framework may be applied on a delay-embedded observed state.

The steady-state score is learned directly from samples using denoising score matching~\cite{Hyvarinen2005,Vincent2011,giorgini2025kgmm}, leveraging the scalability of modern score-based generative modeling~\cite{SongSohlDicksteinKingma2021,Giorgini2024PRLScore,giorgini2025predicting, giorgini2025statistical}. The remaining constant coefficients are obtained from short-lag coordinate-correlation identities evaluated via a clustering-based finite-volume discretization on a data-adaptive partition of state space, leveraging recent advances in scalable clustering and operator inference for high-dimensional dynamical systems~\cite{falasca2024data,falasca2025FDT,giorgini2025learning,giorgini2024reduced,souza2024representing_a,souza2024representing_b,souza2024modified}; the resulting SDE enables efficient sampling of the steady-state distribution and generation of synthetic trajectories satisfying the imposed short-lag dynamical constraints, without direct simulation of the underlying full dynamics.

We validate the approach on two low-dimensional stochastic systems and on a partially-observed spatiotemporally chaotic Kuramoto--Sivashinsky example. In all cases, the learned Markov surrogate reproduces the invariant distribution of the resolved variables and matches their two-time correlation functions.

\textit{Method.---}We consider the Langevin SDE with multiplicative noise
\begin{equation}
    \dot{\vvec x}=\vvec F(\vvec x)+\sqrt{2} \bm\Sigma(\vvec x)\,\vvec\xi(t),
    \label{eq:sde_intro}
\end{equation}
where $\vvec{x}(t) \in \mathbb{R}^D$ is the state vector, $\vvec{F}: \mathbb{R}^D \to \mathbb{R}^D$ is the deterministic drift, $\bm{\Sigma}(\vvec x) \in \mathbb{R}^{D\times D}$ is a state-dependent noise-amplitude matrix, and $\vvec{\xi}(t)$ represents Gaussian white noise with zero mean and unit covariance. We define the diffusion tensor $\bm{D}(\vvec{x}) = \bm{\Sigma}(\vvec{x})\bm{\Sigma}(\vvec{x})^T$, assumed positive definite. This tensor enters the fluctuation--dissipation theorem and relates the spontaneous fluctuations of the system to its linear response~\cite{gomes2025fluctuation,fyodorov2025nonorthogonal}.

The evolution of the probability density $p(\vvec{x},t)$ is governed by the Fokker--Planck equation
\begin{equation}
\label{eq:fokker_planck}
\begin{aligned}
\frac{\partial p}{\partial t}
=&\; -\vvec{\nabla} \cdot \left[\vvec{F}(\vvec{x})\,p\right] \\
&\; + \vvec{\nabla} \cdot \left[(\vvec{\nabla}\cdot\bm{D}(\vvec{x}))\,p + \bm{D}(\vvec{x})\,\vvec{\nabla} p\right].
\end{aligned}
\end{equation}

Assuming the existence of a smooth steady-state distribution $p_{\mathrm{ss}}(\vvec{x})$, the stationary condition $\partial_t p_{\mathrm{ss}} = 0$ yields
\begin{equation}
\vvec{\nabla} \cdot \left[\left(\vvec{F}(\vvec{x}) - \bm{D}(\vvec{x})\vvec{\nabla} \ln p_{\mathrm{ss}}(\vvec{x}) - \vvec{\nabla}\cdot\bm{D}(\vvec{x})\right)p_{\mathrm{ss}}(\vvec{x})\right] = 0.
\end{equation}

This suggests that the drift $\vvec{F}(\vvec{x})$ can be decomposed as
\begin{equation}
\vvec{F}(\vvec{x}) = \bm{D}(\vvec{x})\vvec{\nabla} \ln p_{\mathrm{ss}}(\vvec{x}) + \vvec{\nabla}\cdot\bm{D}(\vvec{x}) + \vvec{g}(\vvec{x}),
\end{equation}
where $(\vvec{\nabla}\cdot\bm{D})_i = \sum_j \partial_j D_{ij}$, the first two terms constitute the conservative (time-reversible) component, and $\vvec{g}(\vvec{x})$ is the non-conservative (time-irreversible) component. We also denote the steady-state score by $\vvec{s}(\vvec{x})=\vvec{\nabla}\ln p_{\mathrm{ss}}(\vvec{x})$.

The non-conservative term $\vvec{g}(\vvec{x})$ satisfies
\begin{equation}
\label{eq:g_constraint}
\vvec{\nabla} \cdot \vvec{g}(\vvec{x}) + \vvec{g}(\vvec{x})\cdot\vvec{\nabla}\ln p_{\mathrm{ss}}(\vvec{x}) = 0.
\end{equation}

We express $\vvec{g}(\vvec{x})$ in terms of an antisymmetric tensor field $\bm{R}(\vvec{x})$,
\begin{equation}
\label{eq:g_R_representation}
 \vvec{g}(\vvec{x}) = \vvec{\nabla} \cdot \bm{R}(\vvec{x}) + \bm{R}(\vvec{x}) \vvec{\nabla} \ln p_{\mathrm{ss}}(\vvec{x}),
 \end{equation}
where $\bm{R}(\vvec{x})^T = -\bm{R}(\vvec{x})$. Conversely, under standard regularity and boundary/decay assumptions on $p_{\mathrm{ss}}$ and $\vvec{g}$, any sufficiently smooth $\vvec{g}$ satisfying Eq.~\eqref{eq:g_constraint} admits a representation of the form~\eqref{eq:g_R_representation} for some antisymmetric $\bm{R}$ (not unique); see Sec.~I of the Supplementary Material~\cite{SM} for a constructive argument.

The full Langevin equation with state-dependent $\bm{D}(\vvec{x})$ and $\bm{R}(\vvec{x})$ reads
\begin{equation}
\label{eq:langevin_full}
\begin{aligned}
\dot{\vvec{x}}(t)
=&\; \bm{D}(\vvec{x})\,\vvec{s}(\vvec{x})
  + \vvec{\nabla} \cdot \bm{D}(\vvec{x})
  + \vvec{\nabla} \cdot \bm{R}(\vvec{x}) \\
&\; + \bm{R}(\vvec{x})\,\vvec{s}(\vvec{x})
  + \sqrt{2}\bm{\Sigma}(\vvec{x})\,\vvec{\xi}(t).
\end{aligned}
\end{equation}
To derive constraints on $\bm{D}$ and $\bm{R}$ from data, we multiply both sides by $\vvec{x}^T$ and average over the steady state. The noise term vanishes by independence. Applying Stein's identity~\cite{stein1981estimation}, with $\langle \bm{D}(\vvec{x})\,\vvec{s}(\vvec{x})\,\vvec{x}^T \rangle + \langle (\vvec{\nabla}\cdot\bm{D}(\vvec{x}))\,\vvec{x}^T \rangle = -\langle \bm{D}(\vvec{x}) \rangle$ and $\langle \vvec{g}(\vvec{x})\,\vvec{x}^T \rangle = -\langle \bm{R}(\vvec{x}) \rangle$ derived in Sec.~III of the Supplementary Material~\cite{SM}, we obtain
\begin{equation}
\label{eq:Cdot_decomposition}
\dot{\bm{C}}(0^+) = -\langle \bm{D} \rangle - \langle \bm{R} \rangle,
\end{equation}
where $\bm{C}(\tau) = \langle \vvec{x}(t+\tau) \vvec{x}(t)^T \rangle$ is the time-correlation matrix. For diffusion processes $\bm{C}(\tau)$ has a cusp at $\tau=0$ due to quadratic variation; we therefore interpret $\dot{\bm{C}}(0)$ as the right-derivative $\dot{\bm{C}}(0^+)\equiv \lim_{\tau\downarrow 0}(\bm{C}(\tau)-\bm{C}(0))/\tau$. In practice, $\dot{\bm{C}}(0^+)$ is estimated from a finite-volume discretization of the dynamics via the rate matrix $\bm{Q}$ of the discretized Markov process: $\dot{\bm{C}}(0^+)\approx \vvec{X}\bm{Q}\,\mathrm{diag}(\vvec{\pi})\vvec{X}^T$, where $\vvec{X}$ collects cluster centroids and $\vvec{\pi}$ is the stationary distribution (see Sec.~III of the Supplementary Material~\cite{SM}). Decomposing $\dot{\bm{C}}(0^+)$ into symmetric and antisymmetric parts,
\begin{equation}
\label{eq:diffusion_constraint}
\langle \bm{D} \rangle = -\dot{\bm{C}}_S(0^+), \qquad \langle \bm{R} \rangle = -\dot{\bm{C}}_A(0^+),
\end{equation}
where $\dot{\bm{C}}_S = \tfrac{1}{2}(\dot{\bm{C}} + \dot{\bm{C}}^T)$ and $\dot{\bm{C}}_A = \tfrac{1}{2}(\dot{\bm{C}} - \dot{\bm{C}}^T)$. These relations directly link the mean diffusion tensor $\langle \bm{D} \rangle$ to the symmetric part of $\dot{\bm{C}}(0^+)$ and the mean antisymmetric tensor $\langle \bm{R} \rangle$ to its antisymmetric part.

% Because density evolution in the discretized space is linear ($\dot{\vvec{\rho}}=\bm{Q}\vvec{\rho}$), the spectrum of $\bm{Q}$ approximates the relaxation rates of the coarse-grained dynamics, with accuracy improving as the number of control volumes increases. 

Eqs.~\eqref{eq:langevin_full}, \eqref{eq:diffusion_constraint} are the central identities underpinning our construction: they separate the conservative (time-reversible) contribution to the drift, fixed by the diffusion tensor $\bm{D}(\vvec{x})$, from the non-conservative, current-carrying component encoded by the antisymmetric tensor field $\bm{R}(\vvec{x})$. The symmetric part of the short-lag correlation derivative $\dot{\bm{C}}(0^+)$ uniquely determines the mean diffusion tensor $\langle \bm{D} \rangle$, while its antisymmetric part fixes the steady-state average $\langle \bm{R} \rangle$ and hence the mean irreversible circulation. The state dependence of both $\bm{D}(\vvec{x})$ and $\bm{R}(\vvec{x})$ can in principle be tuned to match additional dynamical statistics, provided the constraints on $\langle \bm{D} \rangle$ and $\langle \bm{R} \rangle$ in Eq.~\eqref{eq:diffusion_constraint} are satisfied.

In this work we adopt the mean-field approximations $\bm{D}(\vvec{x}) \approx \langle \bm{D} \rangle$ and $\bm{R}(\vvec{x}) \approx \langle \bm{R} \rangle$, under which $\vvec{\nabla} \cdot \bm{D}$ and $\vvec{\nabla} \cdot \bm{R}$ vanish. This closure matches only the \emph{mean} diffusion and antisymmetric components (unconditional circulation) and does not reconstruct state-dependent probability currents; allowing $\bm{D}(\vvec{x})$ and $\bm{R}(\vvec{x})$ to vary spatially is a natural extension left for future work. Under these approximations we obtain the reduced Langevin equation
\begin{equation}
\label{eq:langevin_phi}
\dot{\vvec{x}}(t) = \bm{\Phi}\,\vvec{\nabla} \ln p_{\mathrm{ss}}(\vvec{x}) + \sqrt{2}\bm{\Sigma}\,\vvec{\xi}(t),
\end{equation}
where the drift matrix $\bm{\Phi} = \bm{\Phi}_S + \bm{\Phi}_A$ satisfies
\begin{equation}
\label{eq:phi_from_correlation}
\bm{\Phi} = -\dot{\bm{C}}(0^+),
\end{equation}
with $\bm{\Phi}_S = \langle \bm{D} \rangle = \bm{\Sigma}\bm{\Sigma}^T$ and $\bm{\Phi}_A = \langle \bm{R} \rangle$. The noise-amplitude matrix $\bm{\Sigma}$ is obtained via Cholesky decomposition of $\bm{\Phi}_S$.

The construction therefore proceeds in two steps. First, Eq.~\eqref{eq:langevin_full} defines a family of Langevin equations for which \(p_{\mathrm{ss}}\) is stationary by construction, independently of the particular admissible choices of the positive definite diffusion tensor \(\bm D(\vvec{x})\) and antisymmetric tensor \(\bm R(\vvec{x})\). Second, we select the simplest constant closure, \(\bm D=\langle \bm D\rangle\) and \(\bm R=\langle \bm R\rangle\), with values fixed by the observed short-lag correlation constraint \eqref{eq:diffusion_constraint}. In general, it does not enforce all finite-lag correlations or higher-order path statistics; finite-lag coordinate correlations are recovered by the same closure only under stronger assumptions, for example when \(\mathbb{E}[\vvec{x}(t)\mid \vvec{x}(0)=\vvec{x}_0]\) is affine in \(\vvec{x}_0\), whereas more general cases require state-dependent corrections~\cite{giorgini2026conditional}.

For score-function estimation we train a neural network using the denoising score matching (DSM) loss~\cite{Vincent2011,SongSohlDicksteinKingma2021}, which provides a scalable method for learning the score directly from trajectory data. DSM at fixed noise level $\sigma$ learns the score $\vvec{s}_\sigma$ of the perturbed density $p_\sigma = p_{\mathrm{ss}}\ast \mathcal{N}(0,\sigma^2\bm{I})$; the constructed SDE therefore preserves $p_\sigma$ exactly and approaches $p_{\mathrm{ss}}$ as $\sigma\to 0$. See Sec.~II of the Supplementary Material~\cite{SM} for details on the DSM loss and its connection to Gaussian mixture models. In practice, when the learned score $\vvec{s}_\sigma$ is evaluated at finite noise level $\sigma$, we use the estimator $\bm{\Phi}=\dot{\bm{C}}(0^+)\bm{V}^{-1}$ with $\bm{V}=\langle \vvec{s}_\sigma(\vvec{y})\vvec{y}^T\rangle_{\vvec{y}\sim p_\sigma}$; when the score is accurate, $\bm{V}\approx -\bm{I}$, recovering $\bm{\Phi}\approx -\dot{\bm{C}}(0^+)$.

\textit{Results.---}We first validated the framework on two canonical stochastic systems: a one-dimensional nonlinear SDE with multiplicative noise and a two-dimensional asymmetric four-well potential with a non-conservative rotational component in the drift. In both cases the reduced Langevin model accurately reproduces both the stationary distributions and autocorrelation functions of the original dynamics; full details appear in Sec.~VI of the Supplementary Material~\cite{SM}.

We now apply the framework to the Kuramoto--Sivashinsky (KS) equation, a prototypical model of spatiotemporal chaos arising in pattern formation, flame-front dynamics, and fluid instabilities~\cite{kuramoto1978diffusion,sivashinsky1977nonlinear}. The one-dimensional KS equation on a periodic domain is
\begin{equation}
\label{eq:KS}
\frac{\partial u}{\partial t} = -\Delta u - \Delta^2 u - \frac{1}{2}\nabla\!\left(u^2\right),
\end{equation}
where $u(x,t)$ is a scalar field, $\Delta = \partial^2/\partial x^2$ is the Laplacian, $\nabla = \partial/\partial x$ is the spatial derivative, and the nonlinear term represents advection. The domain size $L$ controls the transition to chaotic dynamics. The KS equation exhibits high-dimensional chaotic attractors and has been extensively studied as a benchmark for reduced-order modeling and data-driven methods~\cite{CvitanovicDavidchackSiminos2010, LuLinChorin2017}.

We simulate Eq.~\eqref{eq:KS} on a periodic domain with $L=34$ using a spectral method with $n_{\text{grid}}=512$ Fourier modes. Subsampling with stride $n_{\text{stride}}=32$ yields reduced state vectors of dimension $D=32$. The trajectory data consists of $10^6$ snapshots with sampling interval $\Delta t = 1$; since the decorrelation time is approximately $50$ time units, this corresponds to roughly $2\times 10^4$ effectively independent samples. Crucially, the resolved state $\vvec{x}(t)\in\mathbb{R}^D$ is a spatially subsampled representation of the full KS field; the unobserved grid points act as hidden variables whose effect on the resolved coordinates is not closed. We apply the framework of Eqs.~\eqref{eq:langevin_phi}--\eqref{eq:phi_from_correlation} to learn the score of the marginal steady-state density of the resolved variables and to construct the drift matrix $\bm{\Phi}$.
%The motivation for modeling this deterministic system with a stochastic surrogate is twofold. First, since the dynamics are only partially observed, the induced evolution on the resolved coordinates $\vvec{x}$ is not closed and is generically non-Markovian; within a Markovian closure, the influence of unresolved degrees of freedom manifests as effective stochasticity. Second, the chaotic nature of KS dynamics introduces exponential sensitivity that restricts pathwise predictability to a Lyapunov time; validation beyond that horizon is therefore necessarily statistical (invariant measures and correlation functions) rather than trajectory shadowing~\cite{Ott2002}.

%Our aim is precisely such a statistically faithful stochastic surrogate for the resolved KS degrees of freedom, in line with earlier data-driven stochastic reductions of KS~\cite{LuLinChorin2017}.  

\begin{figure*}[htbp]
  \centering
  \includegraphics[width=\textwidth]{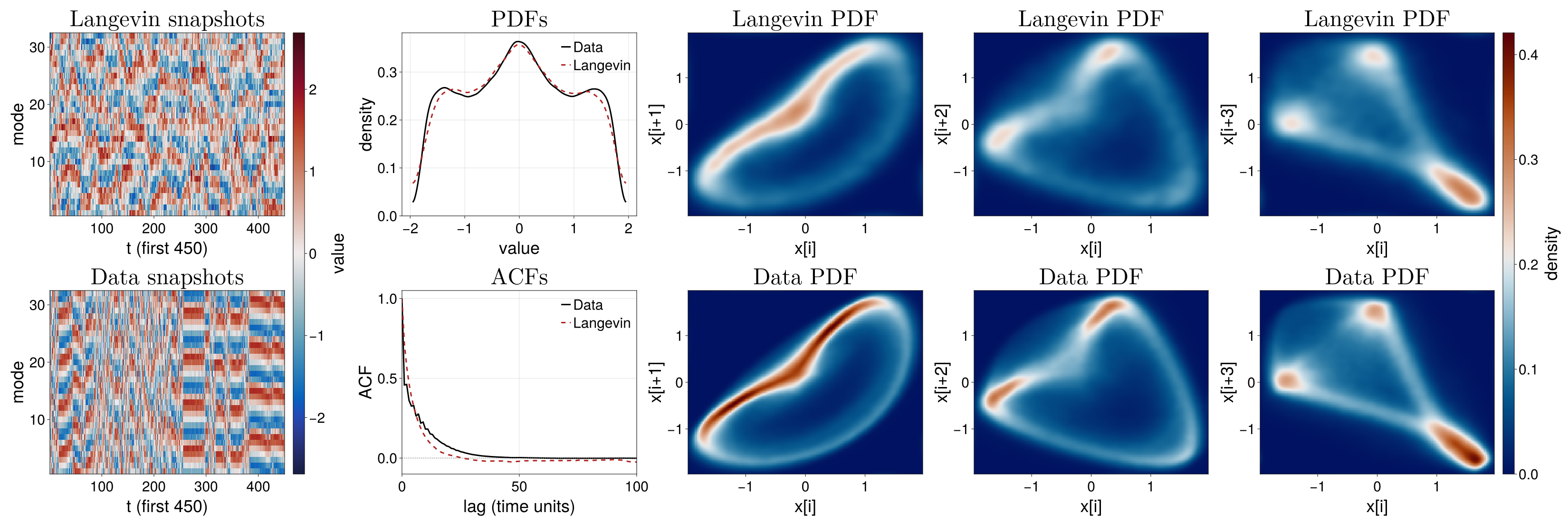}
  \caption{\textbf{Kuramoto--Sivashinsky reduced model validation.} \textbf{Left:} Spatiotemporal evolution (Hovm{\"o}ller plots) of the spectral modes for the reduced Langevin model (top) and the ground truth data (bottom). \textbf{Center:} Marginal invariant distribution (PDF) and autocorrelation function (ACF) for a representative mode (all modes are statistically equivalent due to periodic boundary conditions), comparing the model (red) with data (blue). \textbf{Right:} Joint probability densities $p(x_i, x_{i+k})$ for lags $k=1$, $k=2$, and $k=3$ (left to right), showing the ability of the model to capture spatial correlations between modes.}
  \label{fig:KS_trajectories}
\end{figure*}

Fig.~\ref{fig:KS_trajectories} compares trajectories, marginal probability density functions (PDFs), and autocorrelation functions (ACFs) between the original KS dynamics and our reduced Langevin model. The reduced model accurately reproduces the long-time statistical structure: PDFs (both univariate and bivariate) exhibit near-perfect overlap with the empirical distributions, confirming preservation of the invariant measure, and ACFs are accurately matched over the correlation timescales of the resolved modes. At the level of individual realizations the KS flow displays spatiotemporal chaos organized around unstable coherent structures (including traveling waves and modulated traveling waves), with intermittent transitions and symmetry-related drift episodes~\cite{CvitanovicDavidchackSiminos2010,LanCvitanovic2008}. Because our construction yields a Markovian diffusion process, it is not expected to reproduce this fine-scale intermittency or to shadow a specific KS trajectory at short times.

\begin{figure*}[htbp]
  \centering
  \includegraphics[width=\textwidth]{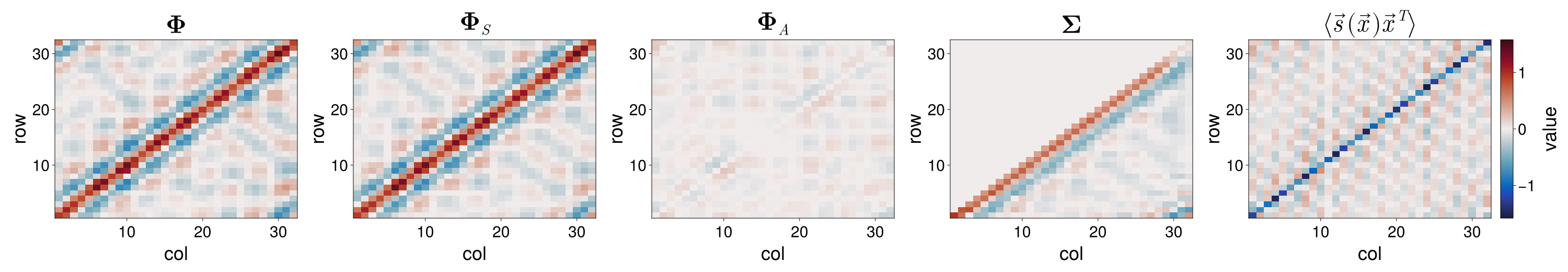}
  \caption{\textbf{Learned operators for the KS equation.} Comparison of the learned linear operators. From left to right: the full drift matrix $\bm{\Phi}$, the symmetric part $\bm{\Phi}_S$ (determining the diffusion tensor), the antisymmetric part $\bm{\Phi}_A$, the noise amplitude matrix $\bm{\Sigma}$ (lower triangular), and the score-position correlation matrix $\langle \vvec{s}\,\vvec{x}^T \rangle$ verifying Stein's identity ($\approx -\bm{I}$).}
  \label{fig:operators}
\end{figure*}

The decomposition of $\bm{\Phi}$ into symmetric and antisymmetric parts reveals the physical structure of the reduced dynamics (Fig.~\ref{fig:operators}). The symmetric component $\bm{\Phi}_S$, which determines the diffusion tensor $\bm{\Sigma}\bm{\Sigma}^T$, captures dissipative processes driving the system toward steady state. Notably, $\bm{\Phi}_A \approx 0$, reflecting the specific $O(2)$ symmetry of KS on a periodic domain: the invariance under $u(x)\mapsto -u(-x)$ makes left- and right-traveling waves statistically equivalent. For PDEs whose reduced dynamics has a preferred orientation in phase space that is not removed by a symmetry of the invariant measure, the same construction yields a nonzero antisymmetric component; for instance, adding a linear odd-derivative dispersive term such as $\gamma\partial_x^3 u$ to KS breaks the reflection symmetry responsible for the cancellation. In this example, while individual KS realizations exhibit pronounced irreversibility through traveling waves, the constant-$\bm{\Phi}$ closure captures only the symmetry-allowed mean antisymmetric component; state-dependent irreversible currents remain beyond this closure (see Sec.~IV of the Supplementary Material~\cite{SM}). The rightmost panel displays $\langle \vvec{s}(\vvec{x}),\vvec{x}^T \rangle \approx -\bm{I}$, computed on perturbed samples $\vvec{y}\sim p*\sigma$, providing numerical verification of Stein's identity and confirming the accuracy of the learned score.

\textit{Conclusions.---}We have presented a data-driven framework for constructing Markovian Langevin equations that bypasses the iterative calibration typically required to match target statistics. By parameterizing the drift in terms of a learned score function and inferring the mobility matrix from short-lag correlation identities, our construction guarantees the steady-state distribution \emph{by design} and enforces the short-lag coordinate-correlation constraints directly from data---so that the surrogate satisfies the targeted invariant measure and these correlation constraints from the outset, without trial-and-error tuning. Combining denoising score matching for learning the score function with finite-volume discretization for estimating time-correlation derivatives, our method identifies a drift matrix $\bm{\Phi}$ whose symmetric part determines the diffusion tensor and whose antisymmetric part captures mean nonequilibrium probability currents.

Application to the Kuramoto--Sivashinsky equation demonstrates the method's capability for spatiotemporally chaotic systems: from trajectory data alone we reconstruct a 32-dimensional Langevin model whose marginal distributions and autocorrelation functions closely match the original PDE dynamics. The decomposition reveals $\bm{\Phi}_A \approx 0$, reflecting the reflection symmetry of KS that causes left- and right-traveling contributions to cancel.

The framework addresses the calibration bottleneck that limits existing techniques. Unlike methods with restrictive parametric assumptions---such as linear inverse models, which impose Gaussian steady states---our approach places no constraint on the form of the invariant measure: the learned score captures the full non-Gaussian structure of the data. At the same time, unlike approaches that require iterative calibration-by-simulation, our construction guarantees the invariant measure \emph{by construction} and enforces the imposed short-lag coordinate-correlation constraints directly from data, so that no trial-and-error tuning is needed. Geometrically, the method decouples density (learned via score matching) from kinetics (inferred via correlation identities), voiding the identifiability and numerical issues associated with joint drift–diffusion inference. It requires no numerical time derivatives of trajectories; instead, we estimate an effective generator on a data-adaptive partition and compute the needed identities from that coarse rate matrix. Compared with sparse-regression or library-based approaches that require specifying a dictionary of candidate basis functions for the drift, our method makes no such ansatz; the drift is expressed in terms of the learned score and a constant matrix inferred from correlation data. Finally, the explicit conservative--non-conservative decomposition ($\bm{\Phi}=\bm{\Phi}_S+\bm{\Phi}_A$) provides physical interpretability, distinguishing dissipative processes driving the system toward steady state from those maintaining nonequilibrium circulation. The main limitations of the present constant-$\bm{\Phi}$ approach are that it reconstructs the invariant density and the imposed short-lag coordinate-correlation constraints, but does not in general enforce broader dynamical features such as finite-lag transition laws, higher-order multi-time observables, or pathwise recovery of individual trajectories.

Several extensions are immediate. Allowing state-dependent $\bm{D}(\vvec{x})$ and $\bm{R}(\vvec{x})$, and incorporating additional observables, would broaden the class of dynamical constraints that can be imposed while preserving $p_{\mathrm{ss}}$. However, enforcing these additional finite-lag dynamical constraints requires transition-level information: a route in this direction is conditional score-based modeling, which supplements the stationary score $\nabla \ln p_{\mathrm{ss}}$ used here with the conditional score of the finite-lag transition density~\cite{giorgini2026conditional}.
More broadly, the learned generators provide a compact platform for uncertainty quantification and accelerated sampling, and they suggest a route to principled stochastic closures for high-dimensional turbulent and climate systems where partial observability and multiscale effects are intrinsic.

\begin{acknowledgments}
We thank A.~N.~Souza for the KS example and his insightful comments.
\end{acknowledgments}

\textit{Data availability.---}The code and supporting scripts used for the Kuramoto--Sivashinsky calculations are publicly available in the \texttt{ScoreUNet1D.jl} repository~\cite{giorgini2026scoreunet1d}.

% --------------------------------------------
% BIBLIOGRAPHY
% --------------------------------------------
\bibliographystyle{apsrev4-2}
\bibliography{references}

%% -------------------------------------------------
%%  APPENDIX
%% -------------------------------------------------
\clearpage
\onecolumngrid

\begin{center}
{\Large\bfseries Appendix}
\end{center}

% --------------------------------------------
% APPENDIX SECTIONS
% --------------------------------------------
% Force section numbering in revtex4-2 PRL style
\setcounter{secnumdepth}{3}
\setcounter{section}{0}
\renewcommand{\thesection}{\Roman{section}}
\renewcommand{\thesubsection}{\thesection.\Alph{subsection}}
\renewcommand{\thesubsubsection}{\thesubsection.\arabic{subsubsection}}

% Override revtex section formatting to include numbers
\makeatletter
\renewcommand\section{\@startsection{section}{1}{\z@}%
                                   {-3.5ex \@plus -1ex \@minus -.2ex}%
                                   {2.3ex \@plus.2ex}%
                                   {\normalfont\Large\bfseries}}
\renewcommand\subsection{\@startsection{subsection}{2}{\z@}%
                                     {-3.25ex\@plus -1ex \@minus -.2ex}%
                                     {1.5ex \@plus .2ex}%
                                     {\normalfont\large\bfseries}}
\renewcommand\subsubsection{\@startsection{subsubsection}{3}{\z@}%
                                     {-3.25ex\@plus -1ex \@minus -.2ex}%
                                     {1.5ex \@plus .2ex}%
                                     {\normalfont\normalsize\bfseries}}
\makeatother

Throughout the Supplementary Material we denote the steady-state density by $p_{\mathrm{ss}}(\vvec{x})$ and write $\vvec{s}(\vvec{x})=\vvec{\nabla}\ln p_{\mathrm{ss}}(\vvec{x})$ for its score.

\section{Verification of the Decomposition of \texorpdfstring{$\vvec{g}(\vvec{x})$}{g(x)}}
\label{appendix:decomposition_proof}

In this section we establish the relationship $\vvec{g}(\vvec{x})=\vvec{\nabla}\cdot\bm{R}(\vvec{x})+\bm{R}(\vvec{x})\,\vvec{s}(\vvec{x})$ stated in the main text, where $\bm{R}(\vvec{x})$ is an antisymmetric tensor field. We prove two complementary results: first, that any drift of this form automatically satisfies the stationarity constraint $\vvec{\nabla}\cdot(\vvec{g}\,p_{\mathrm{ss}})=0$; second, that conversely, any smooth divergence-free probability current admits such a representation (though not uniquely).

\subsection{The Decomposition Enforces Stationarity}
\label{appendix:stationarity_verification}

We first verify that parameterizing the non-conservative drift as $\vvec{g}(\vvec{x})=\vvec{\nabla}\cdot\bm{R}(\vvec{x})+\bm{R}(\vvec{x})\,\vvec{s}(\vvec{x})$, with $\bm{R}(\vvec{x})$ antisymmetric, automatically enforces stationarity. Substituting this form gives
\begin{equation}
\begin{aligned}
\vvec{g}(\vvec{x})\,p_{\mathrm{ss}}(\vvec{x})
&= (\vvec{\nabla} \cdot \bm{R})(\vvec{x})\,p_{\mathrm{ss}}(\vvec{x}) + \bm{R}(\vvec{x})\,\vvec{\nabla} p_{\mathrm{ss}}(\vvec{x}) \\
&= \vvec{\nabla} \cdot \bigl(\bm{R}(\vvec{x})\,p_{\mathrm{ss}}(\vvec{x})\bigr),
\end{aligned}
\end{equation}
where in components $\bigl(\vvec{\nabla} \cdot (\bm{R}p_{\mathrm{ss}})\bigr)_i = \sum_j \partial_j (R_{ij} p_{\mathrm{ss}})$. Taking the divergence once more,
\begin{equation}
\vvec{\nabla} \cdot (\vvec{g} p_{\mathrm{ss}}) = \sum_{i,j} \partial_i \partial_j (R_{ij} p_{\mathrm{ss}})=0,
\end{equation}
since $R_{ij} p_{\mathrm{ss}} = -R_{ji} p_{\mathrm{ss}}$ is antisymmetric and mixed partial derivatives commute. In particular, if $\bm{R}$ is constant then $\vvec{\nabla}\cdot\bm{R}=0$ and $\vvec{g} = \bm{R}\,\vvec{s}$ satisfies the constraint identically.

\subsection{Existence of the Antisymmetric Tensor and Gauge Freedom}
\label{appendix:R_existence}

We now prove the converse: given a smooth $\vvec{g}$ satisfying $\vvec{\nabla}\cdot(\vvec{g}\,p_{\mathrm{ss}})=0$, one can construct an antisymmetric $\bm{R}$ such that $\vvec{g}=\vvec{\nabla}\cdot\bm{R}+\bm{R}\,\vvec{s}$ holds (under mild regularity and decay assumptions). The construction also reveals that the representation is not unique---a manifestation of gauge freedom.

Let $\vvec{J}(\vvec{x}) := \vvec{g}(\vvec{x})\,p_{\mathrm{ss}}(\vvec{x})$ denote the stationary probability current. The constraint $\vvec{\nabla}\cdot(\vvec{g}\,p_{\mathrm{ss}})=0$ is equivalent to $\vvec{\nabla}\cdot\vvec{J}=0$, i.e., $\vvec{J}$ is divergence-free. On $\mathbb{R}^d$, assuming $\vvec{J}$ is sufficiently smooth and decays sufficiently fast at infinity, consider the vector Poisson problem
\begin{equation}
-\Delta \vvec{\psi}(\vvec{x}) = \vvec{J}(\vvec{x}), \qquad \vvec{\psi}(\vvec{x})\to \vvec{0}\ \text{as}\ \|\vvec{x}\|\to\infty,
\end{equation}
where $\Delta=\sum_{i=1}^d \partial_i^2$ is the Laplacian. Taking the divergence and using $\vvec{\nabla}\cdot\vvec{J}=0$ yields $-\Delta(\vvec{\nabla}\cdot\vvec{\psi})=0$, and the decay condition implies $\vvec{\nabla}\cdot\vvec{\psi}=0$. Define an antisymmetric tensor field $\bm{A}$ with components
\begin{equation}
A_{ij}(\vvec{x}) := \partial_i \psi_j(\vvec{x}) - \partial_j \psi_i(\vvec{x}),
\end{equation}
so that $\bm{A}^T=-\bm{A}$. A direct computation gives
\begin{equation}
\bigl(\vvec{\nabla}\cdot\bm{A}\bigr)_i
=\sum_j \partial_j A_{ij}
=-\Delta \psi_i + \partial_i(\vvec{\nabla}\cdot\vvec{\psi})
=J_i,
\end{equation}
and therefore $\vvec{J}=\vvec{\nabla}\cdot\bm{A}$. Finally, set $\bm{R}(\vvec{x}) := \bm{A}(\vvec{x})/p_{\mathrm{ss}}(\vvec{x})$, which gives $\vvec{g}\,p_{\mathrm{ss}}=\vvec{\nabla}\cdot(\bm{R}p_{\mathrm{ss}})$ and hence $\vvec{g}=\vvec{\nabla}\cdot\bm{R}+\bm{R}\,\vvec{s}$.

The representation is not unique: if $\bm{B}$ is any antisymmetric tensor field with $\vvec{\nabla}\cdot\bm{B}=\vvec{0}$, then $\bm{A}+\bm{B}$ produces the same current, leading to a family of admissible $\bm{R}$ fields.

\subsection{Divergence-Free Special Case}
\label{appendix:divergence_free}

We briefly note a special case relevant when $\vvec{g}$ itself is divergence-free. The stationarity constraint $\vvec{\nabla}\cdot(\vvec{g}\,p_{\mathrm{ss}})=0$ then implies
\begin{equation}
\vvec{\nabla}\cdot\vvec{g}=0
\quad\Longrightarrow\quad
\vvec{g}\cdot \vvec{s} = 0,
\end{equation}
so that $\vvec{g}$ is everywhere orthogonal to the score. In this case we can write $\vvec{g} = \bm{R}_{df}\,\vvec{s}$ with the antisymmetric tensor field $\bm{R}_{df}$ given explicitly by the wedge formula:
\begin{equation}
\label{eq:wedge_formula}
\bm{R}_{df}(\vvec{x})
=
\frac{1}{\|\vvec{s}(\vvec{x})\|^2}
\left(
\vvec{g}(\vvec{x})\,\vvec{s}(\vvec{x})^{T}
-
\vvec{s}(\vvec{x})\,\vvec{g}(\vvec{x})^{T}
\right).
\end{equation}

\section{Score Function Estimation via Denoising Score Matching}
\label{appendix:score_estimation}

In this section we describe the estimation of the score function $\vvec{s}(\vvec{x})=\vvec{\nabla} \ln p_{\mathrm{ss}}(\vvec{x})$ from data using denoising score matching (DSM)~\cite{Vincent2011,SongSohlDicksteinKingma2021}, the approach employed in the main text to construct the reduced Langevin dynamics.

\subsection{Denoising Score Matching Loss from Gaussian Mixture Models}

Consider approximating the probability density $p(\vvec{x})$ as a Gaussian mixture model (GMM),
\begin{equation}
p(\vvec{x}) = \frac{1}{N} \sum_{i=1}^{N} \mathcal{N}(\vvec{x} \mid \vvec{\mu}_i, \sigma^2 \bm{I}),
\end{equation}
where $\{\vvec{\mu}_i\}_{i=1}^N$ are data points sampled from the steady-state distribution $p_{\mathrm{ss}}(\vvec{x})$ and $\sigma^2$ is the isotropic covariance of the Gaussian kernels. Direct computation of the score function
\begin{equation}
\vvec{\nabla} \ln p(\vvec{x}) = -\frac{1}{\sigma^2} \sum_{i=1}^{N} \frac{\mathcal{N}(\vvec{x} \mid \vvec{\mu}_i, \sigma^2 \bm{I})(\vvec{x} - \vvec{\mu}_i)}{p(\vvec{x})}
\end{equation}
becomes numerically unstable for small $\sigma$, as the density and its gradient become highly sensitive to local fluctuations in the data.

The denoising score matching framework~\cite{Vincent2011,SongSohlDicksteinKingma2021} provides an elegant solution. If $\vvec{x} = \vvec{\mu} + \sigma \vvec{z}$, where $\vvec{z} \sim \mathcal{N}(\vvec{0}, \bm{I})$, the score function can be expressed as
\begin{equation}
\label{eq:dsm_identity}
\vvec{\nabla} \ln p(\vvec{x}) = -\frac{1}{\sigma} \mathbb{E}[\vvec{z} \mid \vvec{x}].
\end{equation}
This identity allows the score function to be computed as the conditional expectation of the noise vector $\vvec{z}$, scaled by $-1/\sigma$.

To train a neural network $\vvec{s}_\theta(\vvec{x})$ to approximate the score function, we minimize the DSM loss
\begin{equation}
\label{eq:dsm_loss}
\mathcal{L}_{\text{DSM}}(\theta) = \mathbb{E}_{\vvec{\mu} \sim p_{\mathrm{ss}}} \mathbb{E}_{\vvec{z} \sim \mathcal{N}(\vvec{0}, \bm{I})} \left\| \vvec{s}_\theta(\vvec{\mu} + \sigma \vvec{z}) + \frac{\vvec{z}}{\sigma} \right\|^2.
\end{equation}
This loss function is derived from the observation that minimizing the expected squared error between the network output and $-\vvec{z}/\sigma$ is equivalent to matching the true score function at the noise level $\sigma$. The optimal network satisfies $\vvec{s}_\theta^*(\vvec{x}) = \vvec{\nabla} \ln p_\sigma(\vvec{x})$, where $p_\sigma$ is the noise-perturbed distribution obtained by convolving the true invariant density $p_{\mathrm{ss}}$ with a Gaussian kernel of width $\sigma$. In the limit $\sigma \to 0$, we recover the score of the true invariant distribution.

\subsection{Direct Evaluation at Cluster Centroids for Low-Dimensional Systems}

For low-dimensional systems (typically $D = \mathcal{O}(10)$), the DSM loss can be evaluated directly at cluster centroids rather than training a neural network end-to-end. This approach, known as the $k$-means Gaussian-mixture method (KGMM)~\cite{giorgini2025kgmm}, provides exact score estimates that serve as training targets for a neural network interpolator.

The procedure is as follows:
\begin{enumerate}
    \item Perturb the original data points $\{\vvec{\mu}_i\}$ by adding Gaussian noise to generate perturbed samples $\vvec{x}_i = \vvec{\mu}_i + \sigma \vvec{z}_i$, where $\vvec{z}_i \sim \mathcal{N}(\vvec{0}, \bm{I})$.
    \item Partition the perturbed samples $\{\vvec{x}_i\}$ into $N_C$ control volumes $\{\Omega_j\}$ using bisecting K-means clustering. Let $\vvec{C}_j$ denote the centroid of cluster $\Omega_j$.
    \item For each cluster $\Omega_j$, compute the conditional expectation of the displacements using Eq.~\eqref{eq:dsm_identity}:
    \begin{equation}
    \mathbb{E}[\vvec{z} \mid \vvec{x} \in \Omega_j] \approx \frac{1}{|\Omega_j|} \sum_{i: \vvec{x}_i \in \Omega_j} \vvec{z}_i.
    \end{equation}
    \item Estimate the score function at the cluster centroid $\vvec{C}_j$:
    \begin{equation}
    \vvec{s}_{\sigma,j}=\vvec{\nabla} \ln p(\vvec{C}_j) \approx -\frac{1}{\sigma} \mathbb{E}[\vvec{z} \mid \vvec{x} \in \Omega_j].
    \end{equation}
    \item Fit a neural network to interpolate the discrete score estimates $\{(\vvec{C}_j, \vvec{s}_{\sigma,j})\}$ across the entire domain.
\end{enumerate}

The number of clusters $N_C$ must be chosen carefully to balance resolution and noise. A practical scaling relation is
\begin{equation}
N_C \propto \sigma^{-D},
\end{equation}
where $D$ is the effective dimensionality of the data. This ensures that clusters remain small enough to capture local gradient structure while containing enough points for robust averaging.

The choice of $\sigma$ is critical. Smaller values yield score estimates closer to the true steady-state distribution but increase statistical noise. Larger values smooth out fluctuations, improving stability but introducing bias. The optimal $\sigma$ balances these competing effects, minimizing bias while maintaining statistical reliability.

\section{Derivation of the Constraint on \texorpdfstring{$\langle \bm{D} \rangle$}{<D>} and \texorpdfstring{$\langle \bm{R} \rangle$}{<R>}}
\label{appendix:drift_matrix}

In this section we derive the fundamental constraint linking the time-correlation derivative $\dot{\bm{C}}(0^+)$ to the mean diffusion tensor $\langle \bm{D} \rangle$ and mean antisymmetric tensor $\langle \bm{R} \rangle$. As discussed in the main text, we consider the general Langevin equation with state-dependent multiplicative noise,
\begin{equation}
\label{eq:langevin_general_SM}
\dot{\vvec{x}}(t) = \bm{D}(\vvec{x}) \vvec{s}(\vvec{x}) + \vvec{\nabla} \cdot \bm{D}(\vvec{x}) + \vvec{\nabla} \cdot \bm{R}(\vvec{x}) + \bm{R}(\vvec{x}) \vvec{s}(\vvec{x}) + \sqrt{2}\bm{\Sigma}(\vvec{x})\,\vvec{\xi}(t),
\end{equation}
where $\bm{D}(\vvec{x}) = \bm{\Sigma}(\vvec{x})\bm{\Sigma}(\vvec{x})^T$ is the diffusion tensor, $\bm{R}(\vvec{x})$ is an antisymmetric tensor field encoding non-conservative currents, and $\vvec{s}(\vvec{x}) = \vvec{\nabla} \ln p_{\mathrm{ss}}(\vvec{x})$ is the score function.

To derive constraints on $\bm{D}$ and $\bm{R}$ from data, we multiply both sides of Eq.~\eqref{eq:langevin_general_SM} by $\vvec{x}^T$ and average over the steady-state distribution $p_{\mathrm{ss}}$. This yields
\begin{equation}
\label{eq:Cdot_derivation_SM}
\dot{\bm{C}}(0^+) = \langle \bm{D}(\vvec{x}) \vvec{s}(\vvec{x}) \vvec{x}^T \rangle + \langle (\vvec{\nabla} \cdot \bm{D}) \vvec{x}^T \rangle + \langle \vvec{g}(\vvec{x}) \vvec{x}^T \rangle,
\end{equation}
where $\vvec{g}(\vvec{x}) = \vvec{\nabla} \cdot \bm{R}(\vvec{x}) + \bm{R}(\vvec{x}) \vvec{s}(\vvec{x})$ is the non-conservative drift component, and the noise term vanishes by independence. We now evaluate each term in turn.

\subsection{Stein's Identity}
\label{appendix:stein_identity}

For a smooth probability density $p(\vvec{x})$ that decays sufficiently fast at infinity, Stein's identity~\cite{stein1981estimation} in component form states that for any smooth function $\phi$ satisfying $\lim_{\|\vvec{x}\|\to\infty} p(\vvec{x})\phi(\vvec{x}) = 0$,
\begin{equation}
\label{eq:stein_component}
\langle s_j(\vvec{x}) \phi(\vvec{x}) \rangle = -\langle \partial_j \phi(\vvec{x}) \rangle,
\end{equation}
where the expectation is taken with respect to $p(\vvec{x})$ and $\vvec{s}(\vvec{x}) = \vvec{\nabla} \ln p(\vvec{x})$ is the score function. Setting $\phi(\vvec{x}) = x_k$ immediately yields
\begin{equation}
\label{eq:stein}
\langle \vvec{s}(\vvec{x}) \vvec{x}^T \rangle_p = -\bm{I}.
\end{equation}

\subsection{Contribution from the Diffusion Tensor \texorpdfstring{$\bm{D}(\vvec{x})$}{D(x)}}
\label{appendix:D_contribution}

We first evaluate $\langle \bm{D}(\vvec{x}) \vvec{s}(\vvec{x}) \vvec{x}^T \rangle$. Using Stein's identity~\eqref{eq:stein_component} with $\phi(\vvec{x}) = D_{ij}(\vvec{x}) x_k$, we obtain
\begin{equation}
\begin{aligned}
\langle s_j(\vvec{x}) D_{ij}(\vvec{x}) x_k \rangle &= -\langle \partial_j (D_{ij}(\vvec{x}) x_k) \rangle \\
&= -\langle (\partial_j D_{ij})(\vvec{x}) x_k \rangle - \langle D_{ik}(\vvec{x}) \rangle,
\end{aligned}
\end{equation}
where we used $\partial_j x_k = \delta_{jk}$. Summing over $j$ and noting that $(\bm{D} \vvec{s})_i = \sum_j D_{ij} s_j$, we find
\begin{equation}
\label{eq:Ds_xT_relation}
\langle (\bm{D}(\vvec{x}) \vvec{s}(\vvec{x}))_i x_k \rangle = -\langle (\vvec{\nabla} \cdot \bm{D})_i x_k \rangle - \langle D_{ik}(\vvec{x}) \rangle,
\end{equation}
which in matrix form reads
\begin{equation}
\label{eq:Ds_xT_matrix}
\langle \bm{D}(\vvec{x}) \vvec{s}(\vvec{x}) \vvec{x}^T \rangle = -\langle (\vvec{\nabla} \cdot \bm{D}) \vvec{x}^T \rangle - \langle \bm{D}(\vvec{x}) \rangle.
\end{equation}

Substituting this into the first two terms of Eq.~\eqref{eq:Cdot_derivation_SM}, we see that the $\langle (\vvec{\nabla} \cdot \bm{D}) \vvec{x}^T \rangle$ contributions cancel, leaving
\begin{equation}
\label{eq:D_contribution_final}
\langle \bm{D}(\vvec{x}) \vvec{s}(\vvec{x}) \vvec{x}^T \rangle + \langle (\vvec{\nabla} \cdot \bm{D}) \vvec{x}^T \rangle = -\langle \bm{D}(\vvec{x}) \rangle.
\end{equation}

\subsection{Contribution from the Antisymmetric Tensor \texorpdfstring{$\bm{R}(\vvec{x})$}{R(x)}}
\label{appendix:R_contribution}

We now evaluate $\langle \vvec{g}(\vvec{x}) \vvec{x}^T \rangle$ where $\vvec{g}(\vvec{x}) = (\vvec{\nabla} \cdot \bm{R})(\vvec{x}) + \bm{R}(\vvec{x}) \vvec{s}(\vvec{x})$ and $(\vvec{\nabla} \cdot \bm{R})_i := \sum_j \partial_j R_{ij}$. The $(i,k)$ entry is
\begin{equation}
\langle g_i(\vvec{x}) x_k \rangle = \Big\langle \sum_j \partial_j R_{ij}(\vvec{x}) x_k \Big\rangle + \Big\langle \sum_j R_{ij}(\vvec{x}) s_j(\vvec{x}) x_k \Big\rangle.
\end{equation}

Applying Stein's identity~\eqref{eq:stein_component} with $\phi(\vvec{x}) = R_{ij}(\vvec{x}) x_k$ for fixed $(i,k)$ and summing over $j$:
\begin{equation}
\begin{aligned}
\Big\langle \sum_j s_j(\vvec{x}) R_{ij}(\vvec{x}) x_k \Big\rangle 
&= -\Big\langle \sum_j \partial_j \big( R_{ij}(\vvec{x}) x_k \big) \Big\rangle \\
&= -\Big\langle \sum_j (\partial_j R_{ij})(\vvec{x}) x_k \Big\rangle - \langle R_{ik}(\vvec{x}) \rangle,
\end{aligned}
\end{equation}
where we used $\partial_j x_k = \delta_{jk}$. Adding the $\langle (\vvec{\nabla} \cdot \bm{R})_i x_k \rangle$ term, the divergence contributions cancel, yielding
\begin{equation}
\label{eq:g_R_relation}
\langle \vvec{g}(\vvec{x}) \vvec{x}^T \rangle = -\langle \bm{R}(\vvec{x}) \rangle.
\end{equation}
Since $\bm{R}(\vvec{x})^T = -\bm{R}(\vvec{x})$, it follows that $\langle \vvec{g}(\vvec{x}) \vvec{x}^T \rangle$ is automatically antisymmetric.

\subsection{Combined Constraint}
\label{appendix:combined_constraint}

Substituting Eqs.~\eqref{eq:D_contribution_final} and~\eqref{eq:g_R_relation} into Eq.~\eqref{eq:Cdot_derivation_SM}, we obtain the central identity
\begin{equation}
\label{eq:Cdot_constraint_SM}
\dot{\bm{C}}(0^+) = -\langle \bm{D}(\vvec{x}) \rangle - \langle \bm{R}(\vvec{x}) \rangle.
\end{equation}
Decomposing into symmetric and antisymmetric parts and noting that $\langle \bm{D} \rangle$ is symmetric while $\langle \bm{R} \rangle$ is antisymmetric, we find
\begin{equation}
\label{eq:DR_constraints_SM}
\langle \bm{D} \rangle = -\dot{\bm{C}}_S(0^+), \qquad \langle \bm{R} \rangle = -\dot{\bm{C}}_A(0^+),
\end{equation}
where $\dot{\bm{C}}_S = \tfrac{1}{2}(\dot{\bm{C}} + \dot{\bm{C}}^T)$ and $\dot{\bm{C}}_A = \tfrac{1}{2}(\dot{\bm{C}} - \dot{\bm{C}}^T)$.

Crucially, the state dependence of both $\bm{D}(\vvec{x})$ and $\bm{R}(\vvec{x})$ can be tuned to match additional dynamical statistics (e.g., higher-order correlations or state-dependent transport), provided the constraints on $\langle \bm{D} \rangle$ and $\langle \bm{R} \rangle$ in Eq.~\eqref{eq:DR_constraints_SM} are satisfied.

\subsection{Mean-Field Approximation}
\label{appendix:mean_field}

In this work we adopt the mean-field approximations
\begin{equation}
\bm{D}(\vvec{x}) \approx \langle \bm{D} \rangle, \qquad \bm{R}(\vvec{x}) \approx \langle \bm{R} \rangle,
\end{equation}
under which the divergence terms $\vvec{\nabla} \cdot \bm{D}$ and $\vvec{\nabla} \cdot \bm{R}$ vanish. This closure matches only the \emph{mean} diffusion and antisymmetric components and does not reconstruct state-dependent probability currents; allowing $\bm{D}(\vvec{x})$ and $\bm{R}(\vvec{x})$ to vary spatially is a natural extension left for future work.

Under these approximations, the Langevin equation~\eqref{eq:langevin_general_SM} reduces to
\begin{equation}
\dot{\vvec{x}}(t) = \bm{\Phi}\,\vvec{s}(\vvec{x}) + \sqrt{2}\bm{\Sigma}\,\vvec{\xi}(t),
\end{equation}
where the drift matrix $\bm{\Phi} = \bm{\Phi}_S + \bm{\Phi}_A$ satisfies $\bm{\Phi} = -\dot{\bm{C}}(0^+)$, with $\bm{\Phi}_S = \langle \bm{D} \rangle = \bm{\Sigma}\bm{\Sigma}^T$ and $\bm{\Phi}_A = \langle \bm{R} \rangle$. The drift matrix is obtained by solving
\begin{equation}
\label{eq:Phi_main}
\bm{\Phi} = \dot{\bm{C}}(0^+) \cdot \langle \vvec{s}(\vvec{x}) \vvec{x}^T \rangle^{-1},
\end{equation}
where $\bm{C}(\tau) = \langle \vvec{x}(t+\tau) \vvec{x}(t)^T \rangle$ is the time-correlation matrix. We now describe how to estimate each of these terms in practice.

\subsubsection{Application to Score Estimation}

A crucial subtlety arises in our framework. The score function learned via DSM is the score of the \emph{perturbed} density $p_\sigma$, not the true invariant density $p_{\mathrm{ss}}$. Therefore, Stein's identity applies when the expectation is taken with respect to the \emph{same} perturbed distribution. Concretely, if $\vvec{s}_\sigma(\vvec{x}) = \vvec{\nabla} \ln p_\sigma(\vvec{x})$, then
\begin{equation}
\langle \vvec{s}_\sigma(\vvec{x}) \vvec{x}^T \rangle_{p_\sigma} = -\bm{I}.
\end{equation}

To estimate this expectation from data, we must sample $\vvec{x}$ from the perturbed distribution $p_\sigma$. Since $p_\sigma$ is obtained by convolving $p_{\mathrm{ss}}$ with a Gaussian kernel of width $\sigma$, we generate samples from $p_\sigma$ by adding Gaussian noise to the original time-series data,
\begin{equation}
\tilde{\vvec{x}}_i = \vvec{x}_i + \sigma \vvec{z}_i, \quad \vvec{z}_i \sim \mathcal{N}(\vvec{0}, \bm{I}),
\end{equation}
where $\{\vvec{x}_i\}$ are the original data points sampled from $p_{\mathrm{ss}}$. The estimator for the score-position correlation matrix is then
\begin{equation}
\langle \vvec{s}_\sigma(\vvec{x}) \vvec{x}^T \rangle_{p_\sigma} \approx \frac{1}{N} \sum_{i=1}^{N} \vvec{s}_\sigma(\tilde{\vvec{x}}_i) \tilde{\vvec{x}}_i^T.
\end{equation}

If instead we evaluate the score at the original (unperturbed) data points, we obtain
\begin{equation}
\langle \vvec{s}_\sigma(\vvec{x}) \vvec{x}^T \rangle_{p_{\mathrm{ss}}} = -\bm{I} + \bm{E}_\sigma,
\end{equation}
where $\bm{E}_\sigma$ is an error term arising from the mismatch between the distributions. This error vanishes in the limit $\sigma \to 0$.

To estimate $\bm{E}_\sigma$ explicitly, note that $\bm{E}_\sigma = \langle (\vvec{s}_\sigma(\vvec{x})-\vvec{s}(\vvec{x}))\vvec{x}^T \rangle_{p_{\mathrm{ss}}}$, where $\vvec{s}=\vvec{\nabla}\ln p_{\mathrm{ss}}$. Since $p_\sigma$ is the Gaussian convolution of $p_{\mathrm{ss}}$, we may write $p_\sigma = e^{(\sigma^2/2)\Delta}p_{\mathrm{ss}}$, with $\Delta$ the Laplacian. For smooth $p_{\mathrm{ss}}$, a small-$\sigma$ expansion yields $\ln p_\sigma = \ln p_{\mathrm{ss}} + \frac{\sigma^2}{2}\bigl(\Delta \ln p_{\mathrm{ss}} + \|\vvec{s}\|^2\bigr) + \mathcal{O}(\sigma^4)$ and therefore
\begin{equation}
\bm{E}_\sigma
=
\frac{\sigma^2}{2}\left\langle \vvec{\nabla}\!\left(\Delta \ln p_{\mathrm{ss}}(\vvec{x}) + \|\vvec{s}(\vvec{x})\|^2\right)\vvec{x}^T \right\rangle_{p_{\mathrm{ss}}}
+ \mathcal{O}(\sigma^4).
\end{equation}
In particular, if $\ln p_{\mathrm{ss}}$ has bounded third derivatives and $\langle \|\vvec{x}\|\rangle_{p_{\mathrm{ss}}}<\infty$, then $\|\bm{E}_\sigma\|=\mathcal{O}(\sigma^2)$.

\subsection{Estimation of \texorpdfstring{$\dot{\bm{C}}(0)$}{Cdot(0)} via Rate Matrix Discretization}

The time derivative of the correlation function at $\tau=0$ can be estimated using a finite-volume discretization of the state space. The key advantage of this approach is that the dynamics of the probability density becomes \emph{linear} in the high-dimensional discretized space, allowing us to compute $\dot{\bm{C}}(0)$ directly from the transition rate matrix.

The state space is partitioned into $N_C$ control volumes $\{\Omega_j\}$, with $\vvec{C}_j$ denoting the centroid of each volume. The evolution of the probability vector $\vvec{\rho}(t)$, where $\rho_j(t)$ represents the probability of the system being in control volume $\Omega_j$ at time $t$, is governed by
\begin{equation}
\dot{\vvec{\rho}} = \bm{Q}\vvec{\rho},
\end{equation}
where $\bm{Q} \in \mathbb{R}^{N_C \times N_C}$ is the rate matrix. The off-diagonal elements $Q_{jk}$ represent the transition rates from volume $k$ to volume $j$, while the diagonal elements are determined by probability conservation:
\begin{equation}
Q_{jj} = -\sum_{k \neq j} Q_{kj}.
\end{equation}
The rate matrix is constructed from empirical transition counts in short-time trajectory data. For details on the construction, see Refs.~\cite{giorgini2024reduced,souza2024representing_a}.

Let $x_i^n$ denote the $i$th component of the centroid of bin $n$, with stationary probability mass $\pi_n$ satisfying $\bm{Q}\vvec{\pi} = \vvec{0}$. The time-correlation matrix can be written as
\begin{equation}
C_{ij}(\tau) = \sum_{n=1}^{N_C} x_j^n \pi_n \sum_{m=1}^{N_C} x_i^m \bigl[e^{\bm{Q}\tau}\bigr]_{mn}.
\end{equation}
Expanding the matrix exponential for small $\tau$,
\begin{equation}
C_{ij}(\tau) \approx \sum_{n=1}^{N_C} x_j^n \pi_n \sum_{m=1}^{N_C} x_i^m \bigl[\bm{I} + \bm{Q}\tau\bigr]_{mn}.
\end{equation}
Taking the derivative at $\tau=0$,
\begin{equation}
\dot{C}_{ij}(0) = \sum_{n,m=1}^{N_C} x_j^n \pi_n \, x_i^m \, Q_{mn}.
\end{equation}

This provides a direct method to compute $\dot{\bm{C}}(0^+)$ from the discretized rate matrix $\bm{Q}$ and the cluster centroids, without requiring numerical differentiation of trajectory data.

\paragraph{Summary of the practical estimator.}
The complete pipeline for estimating $\bm{\Phi}$ is as follows:
\begin{enumerate}
\item Sample $\vvec{y}=\vvec{x}+\sigma \vvec{z}$ (with $\vvec{z}\sim\mathcal{N}(\vvec{0},\bm{I})$) to work under the perturbed density $p_\sigma$;
\item Estimate the Stein matrix $\bm{V}_{\mathrm{data}}\approx \frac{1}{N}\sum_n \vvec{s}_\theta(\vvec{y}_n)\vvec{y}_n^T$;
\item Estimate $\dot{\bm{C}}(0^+)\approx \bm{X}\bm{Q}\,\mathrm{diag}(\vvec{\pi})\bm{X}^T$ from the rate matrix;
\item Solve $\bm{\Phi}\bm{V}_{\mathrm{data}}=\dot{\bm{C}}(0^+)$ for $\bm{\Phi}$.
\end{enumerate}
When the score is accurate, $\bm{V}_{\mathrm{data}}\approx -\bm{I}$ (Stein's identity under $p_\sigma$), recovering $\bm{\Phi}\approx -\dot{\bm{C}}(0^+)$. We report $\bm{V}_{\mathrm{data}}$ as a self-consistency diagnostic in all experiments.

\section{Deterministic limit, coarse-graining-induced diffusion, and symmetry constraints on \texorpdfstring{$\dot{\bm{C}}(0^+)$}{Cdot(0+)}}
\label{sec:det_coarse_sym}

We recall the time-correlation matrix
\begin{equation}
\bm{C}(\tau) \equiv \left\langle \vvec{x}(t+\tau)\vvec{x}(t)^T\right\rangle,
\qquad
\dot{\bm{C}}(0) \equiv \left.\frac{d}{d\tau}\bm{C}(\tau)\right|_{\tau=0},
\end{equation}
and its symmetric/antisymmetric decomposition $\dot{\bm{C}}_S=\tfrac12(\dot{\bm{C}}+\dot{\bm{C}}^T)$ and $\dot{\bm{C}}_A=\tfrac12(\dot{\bm{C}}-\dot{\bm{C}}^T)$. For a stationary process one has $\bm{C}(-\tau)=\bm{C}(\tau)^T$, and therefore, by the definitions of the symmetric and antisymmetric parts, $\bm{C}_S(-\tau)=\bm{C}_S(\tau)$ and $\bm{C}_A(-\tau)=-\bm{C}_A(\tau)$.

\paragraph{Deterministic, fully observed dynamics implies $\dot{\bm{C}}_S(0)=\bm{0}$.}
Assume $\vvec{x}(t)$ evolves deterministically in continuous time,
\begin{equation}
\dot{\vvec{x}} = \vvec{F}(\vvec{x}),
\end{equation}
and that $\bm{C}(\tau)$ is differentiable at $\tau=0$.\footnote{In contrast, for diffusions the right-derivative at $\tau=0$ exists while $\bm{C}(\tau)$ typically exhibits a cusp due to quadratic variation.} For small $\tau>0$,
\begin{equation}
\vvec{x}(t+\tau)=\vvec{x}(t)+\tau\,\vvec{F}(\vvec{x}(t)) + \mathcal O(\tau^2),
\end{equation}
and therefore
\begin{equation}
\bm{C}(\tau)=\bm{C}(0)+\tau\,\left\langle \vvec{F}(\vvec{x})\,\vvec{x}^T\right\rangle + \mathcal O(\tau^2),
\qquad
\dot{\bm{C}}(0)=\left\langle \vvec{F}(\vvec{x})\,\vvec{x}^T\right\rangle.
\label{eq:Cdot_det}
\end{equation}
Stationarity of the second moments implies $0=\frac{d}{dt}\langle \vvec{x}\vvec{x}^T\rangle
=\langle \vvec{F}(\vvec{x})\vvec{x}^T\rangle + \langle \vvec{x}\vvec{F}(\vvec{x})^T\rangle$, hence
\begin{equation}
\dot{\bm{C}}(0)+\dot{\bm{C}}(0)^T=\bm{0}
\qquad\Rightarrow\qquad
\dot{\bm{C}}_S(0)=\bm{0},
\label{eq:CdotS_det_zero}
\end{equation}
so that $\dot{\bm{C}}(0)$ is purely antisymmetric in the deterministic, fully observed limit. In the reduced Langevin representation used in the main text, $\dot{\bm{C}}_S(0)=\bm{0}$ corresponds to $\bm{\Sigma}=\bm{0}$.

\paragraph{Coarse-graining and finite sampling generically yield $\dot{\bm{C}}_S(0)\neq \bm{0}$ through an effective diffusion.}
Consider now a reduced observable $\vvec{y}=\Pi(\vvec{x})$ (e.g., projection onto a subset of modes, POD coordinates, or cluster centroids). Even if the underlying $\vvec{x}(t)$ is deterministic, the reduced increments $\Delta \vvec{y}=\vvec{y}(t+\Delta t)-\vvec{y}(t)$ typically exhibit nontrivial conditional variability because many microstates $\vvec{x}$ correspond to the same reduced state $\vvec{y}$. Moreover, even in the absence of an explicit state-space projection, observing a chaotic system at a finite sampling interval $\Delta t$ induces an effective conditional dispersion of the increments at fixed $\vvec{y}(t)$, due to sensitive dependence on initial conditions and unresolved sub-$\Delta t$ variability. As a result, short-lag estimates of $\dot{\bm{C}}_S(0)$ obtained from discrete-time data can exhibit a nonzero symmetric component, which vanishes only in the joint limit of full observability and $\Delta t\to 0$. A standard small-$\Delta t$ closure is provided by the first two Kramers--Moyal coefficients,
\begin{equation}
\vvec{a}(\vvec{y}) = \lim_{\Delta t\to 0}\frac{1}{\Delta t}\,\mathbb{E}\!\left[\Delta \vvec{y}\,|\,\vvec{y}(t)=\vvec{y}\right],
\qquad
\bm{B}(\vvec{y}) = \lim_{\Delta t\to 0}\frac{1}{2\Delta t}\,\mathbb{E}\!\left[\Delta \vvec{y}\,\Delta \vvec{y}^T\,|\,\vvec{y}(t)=\vvec{y}\right],
\label{eq:KM}
\end{equation}
which motivate the Markov diffusion approximation
\begin{equation}
d\vvec{y} = \vvec{a}(\vvec{y})\,dt + \sqrt{2}\,\bm{\sigma}(\vvec{y})\,d\vvec{W}_t,
\qquad
\bm{\sigma}(\vvec{y})\bm{\sigma}(\vvec{y})^T=\bm{B}(\vvec{y}).
\label{eq:effSDE}
\end{equation}
Applying It\^o's formula to $\vvec{y}\vvec{y}^T$ and using stationarity yields
\begin{equation}
\bm{0}
=
\left\langle \vvec{a}(\vvec{y})\,\vvec{y}^T\right\rangle
+
\left\langle \vvec{y}\,\vvec{a}(\vvec{y})^T\right\rangle
+
2\left\langle \bm{B}(\vvec{y})\right\rangle.
\label{eq:stationary_yy}
\end{equation}
Moreover, the right-derivative of the correlation at $\tau=0$ is $\dot{\bm{C}}_{\!y}(0^+)=\langle \vvec{a}(\vvec{y})\,\vvec{y}^T\rangle$, so taking the symmetric part in \eqref{eq:stationary_yy} gives
\begin{equation}
\dot{\bm{C}}_{\!y,S}(0^+) = -\left\langle \bm{B}(\vvec{y})\right\rangle,
\label{eq:CdotS_effdiff}
\end{equation}
which is generically nonzero under coarse-graining. In the constant-diffusion closure adopted in the main text, $\bm{B}(\vvec{y})\approx \bm{\Sigma}\bm{\Sigma}^T$ and \eqref{eq:CdotS_effdiff} reduces to $\dot{\bm{C}}_S(0^+)=-\bm{\Sigma}\bm{\Sigma}^T$. Equation~\eqref{eq:CdotS_effdiff} thus formalizes the interpretation of $-\dot{\bm{C}}_S(0^+)$ as the total (intrinsic or effective) diffusion required by the reduced Markov description.

\paragraph{Symmetry constraints on $\dot{\bm{C}}(0)$: the cases $SO(2)$ and $O(2)$.}
Let a symmetry group $G$ act linearly on the reduced coordinates via orthogonal matrices $\bm{U}(g)$, $g\in G$. If the dynamics is equivariant and the stationary statistics are $G$-invariant, then for all $\tau$,
\begin{equation}
\bm{C}(\tau)=\bm{U}(g)\,\bm{C}(\tau)\,\bm{U}(g)^T,
\qquad \forall g\in G,
\label{eq:C_sym_constraint}
\end{equation}
and likewise $\dot{\bm{C}}(0)=\bm{U}(g)\dot{\bm{C}}(0)\bm{U}(g)^T$. In a basis adapted to the irreducible representations (irreps) of $G$, these constraints restrict the admissible block structure of $\dot{\bm{C}}(0)$.

A particularly relevant situation for spatially periodic systems is the continuous rotation group $SO(2)$ (e.g., translations on a ring), which acts in each two-dimensional irrep as $\bm{U}(\theta)=\bm{R}(\theta)$ with $\bm{R}(\theta)$ a planar rotation matrix. In such a $2\times2$ irrep, the commutation constraint in \eqref{eq:C_sym_constraint} implies that any admissible block of $\dot{\bm{C}}(0)$ must be of the form
\begin{equation}
\dot{\bm{C}}_k(0)=\alpha_k\,\bm{I}+\beta_k\,\bm{J},
\qquad
\bm{J}=\begin{pmatrix}0&-1\\1&0\end{pmatrix},
\label{eq:SO2_commutant}
\end{equation}
where $\alpha_k$ contributes to the symmetric part and $\beta_k$ to the antisymmetric part. For a fully deterministic, fully observed dynamics, \eqref{eq:CdotS_det_zero} forces $\alpha_k=0$, so that $SO(2)$ symmetry alone still allows a nontrivial antisymmetric block proportional to $\bm{J}$.

If, however, the symmetry group is enlarged to $O(2)$ by including a reflection $\bm{S}$ (with $\det\bm{S}=-1$), then in each $2\times2$ irrep one has $\bm{S}\bm{J}\bm{S}^T=-\bm{J}$. Imposing invariance under reflections in \eqref{eq:C_sym_constraint} therefore enforces $\beta_k=0$ in \eqref{eq:SO2_commutant}. Consequently, under fully $O(2)$-invariant statistics one obtains $\dot{\bm{C}}_{k,A}(0)=\bm{0}$. This explains why, for systems such as Kuramoto--Sivashinsky on a periodic domain (translation + reflection symmetry), the mean antisymmetric component inferred from $\dot{\bm{C}}(0)$ can be strongly suppressed by symmetry even when the underlying dynamics exhibits pronounced state-dependent circulation: symmetry may force the mean oriented rotation to cancel in the unconditional average.

\section{Score U-Net and estimation of \texorpdfstring{$\bm{\Phi}$ and $\bm{\Sigma}$}{Phi and Sigma} (KS figures)}
\label{sec:ks_unet_phi_sigma}

This section documents the numerical and architectural details needed to reproduce Figs.~1--2 of the main text for the Kuramoto--Sivashinsky (KS) experiment: (i) the convolutional U-Net used to learn the steady-state score function via denoising score matching and (ii) the estimator used to compute the drift and diffusion matrices $\bm{\Phi}$ and $\bm{\Sigma}$ from trajectory data.

\subsection{Reduced state, normalization, and tensor layout}
\label{sec:ks_data_layout}

Let $\vvec{u}(t)\in\mathbb{R}^{D}$ denote the reduced KS state used in the main text ($D=32$ spectral degrees of freedom obtained by subsampling the Fourier representation of the PDE solution; see main text for simulation details). The KS dataset used for the main figures consists of $T\sim 10^6$ samples at a fixed sampling interval $\Delta t$ (reported in the main text). All learning and operator estimation are performed in the \emph{componentwise normalized} coordinates
\begin{equation}
\vvec{x}(t)=\bm{S}^{-1}\bigl(\vvec{u}(t)-\vvec{\mu}\bigr),
\qquad
\bm{S}\equiv \mathrm{diag}(\sigma_1,\dots,\sigma_D),
\qquad
\mu_i=\langle u_i\rangle,\ \sigma_i=\sqrt{\mathrm{Var}(u_i)},
\label{eq:ks_normalization}
\end{equation}
so that each component of $\vvec{x}$ has approximately zero mean and unit variance under the empirical steady state. In the implementation, samples are stored in the tensor layout expected by 1D convolutions,
\begin{equation}
\text{data tensor shape:}\qquad (L, C, B)=(D,1,B),
\end{equation}
where $L$ is the ``spatial'' (mode) index, $C$ is the number of channels, and $B$ is the batch size. When needed, we flatten $(L,C)$ into a single state dimension $D=L\,C$.

If one wishes to express the learned objects in the \emph{original} coordinates $\vvec{u}$, note that the score transforms as
\begin{equation}
\vvec{s}_{u}(\vvec{u})\equiv \vvec{\nabla}_{u}\ln p_{u}(\vvec{u})
\;=\;\bm{S}^{-1}\,\vvec{s}_{x}(\vvec{x}),
\qquad \vvec{x}=\bm{S}^{-1}(\vvec{u}-\vvec{\mu}),
\label{eq:score_change_of_variables}
\end{equation}
and the constant matrices in the reduced Langevin model (defined below) transform as
\begin{equation}
\bm{\Phi}_{u}=\bm{S}\,\bm{\Phi}_{x}\,\bm{S},
\qquad
\bm{\Sigma}_{u}=\bm{S}\,\bm{\Sigma}_{x}.
\label{eq:PhiSigma_change_of_variables}
\end{equation}
In the main figures we report $\vvec{x}(t)$, $\bm{\Phi}_x$, and $\bm{\Sigma}_x$ (normalized units), because this is the coordinate system in which the score network is trained and validated.

\subsection{1D score U-Net architecture}
\label{sec:ks_unet_arch}

We parameterize the noise-prediction network $\vvec{\varepsilon}_{\theta}$ as a one-dimensional U-Net acting on the mode index:
\begin{equation}
\vvec{\varepsilon}_{\theta}:\mathbb{R}^{L\times C}\to\mathbb{R}^{L\times C},
\qquad (L,C)=(D,1),
\end{equation}
and interpret the input vector $\vvec{x}\in\mathbb{R}^{D}$ as a 1D signal of length $L=D$ with one channel. The network is composed of:
\begin{enumerate}
    \item \textbf{Convolutional blocks (ConvBlock).} Each block consists of two 1D convolutions with kernel size $k=5$ and ``same'' padding (length-preserving), each followed by batch normalization and a pointwise nonlinearity:
    \begin{equation}
    \text{ConvBlock}:\quad
    \bm{h}\mapsto \phi\!\left(\mathrm{BN}\!\left(\mathrm{Conv}_2\bigl(\phi(\mathrm{BN}(\mathrm{Conv}_1(\bm{h})))\bigr)\right)\right),
    \end{equation}
    where $\phi$ is the Swish activation $\phi(a)=a\,\sigma(a)=a/(1+e^{-a})$.
    \item \textbf{Encoder (down path).} At level $\ell$, a ConvBlock produces a skip tensor and a strided convolution (kernel 2, stride 2, no padding) downsamples the length by a factor two.
    \item \textbf{Bottleneck.} A ConvBlock at the coarsest resolution expands the channel dimension by a factor two.
    \item \textbf{Decoder (up path).} Each level upsamples by nearest-neighbor interpolation (factor two), concatenates the corresponding encoder skip tensor along the channel dimension, and applies a ConvBlock. When the upsampled length does not match the skip length (due to integer division in the downsampling), we apply a symmetric crop/zero-pad to match dimensions before concatenation.
    \item \textbf{Final projection.} A $1\times1$ convolution maps the final feature tensor back to $C=1$ output channel.
\end{enumerate}

For the KS runs shown in the main text we use \textbf{base width} $16$ and \textbf{channel multipliers} $(1,2,4)$, yielding encoder channel sizes $(16,32,64)$ and a bottleneck width $128$, followed by the symmetric decoder. All convolutions enforce \textbf{periodic boundary conditions} on the length index via circular padding: for a kernel of size $k$ and dilation $d$, we pad by $p_{\mathrm{left}}=\lfloor d(k-1)/2\rfloor$ points on the left and $p_{\mathrm{right}}=d(k-1)-p_{\mathrm{left}}$ points on the right by wrapping the signal endpoints. This is appropriate for reduced KS representations where the retained Fourier modes live on a periodic domain.

\subsection{Denoising score-matching training objective}
\label{sec:ks_dsm_training}

Let $p_{\mathrm{ss}}(\vvec{x})$ denote the empirical steady-state density in normalized coordinates. We train $\vvec{\varepsilon}_{\theta}$ using the (single-noise-level) denoising score matching objective~\cite{Vincent2011,SongSohlDicksteinKingma2021}. Draw $\vvec{x}\sim p_{\mathrm{ss}}$ and $\vvec{z}\sim\mathcal{N}(\vvec{0},\bm{I})$, and form the perturbed sample
\begin{equation}
\vvec{y}=\vvec{x}+\sigma\,\vvec{z},
\label{eq:ks_dsm_perturb}
\end{equation}
so that $\vvec{y}\sim p_{\sigma}=p_{\mathrm{ss}}\ast \mathcal{N}(\vvec{0},\sigma^2\bm{I})$. The network is trained to predict $\vvec{z}$ from $\vvec{y}$ by minimizing the mean-squared error
\begin{equation}
\mathcal{L}_{\mathrm{DSM}}(\theta)
=
\frac{1}{2}\,
\mathbb{E}_{\vvec{x}\sim p_{\mathrm{ss}},\,\vvec{z}\sim\mathcal{N}(0,\bm{I})}
\left[\left\|\vvec{\varepsilon}_{\theta}(\vvec{x}+\sigma\vvec{z})-\vvec{z}\right\|^2\right].
\label{eq:ks_dsm_loss}
\end{equation}
By the denoising identity (Eq.~\eqref{eq:dsm_identity}), the score of the perturbed density is
\begin{equation}
\vvec{s}_{\sigma}(\vvec{y})
\equiv
\vvec{\nabla}_{y}\ln p_{\sigma}(\vvec{y})
\approx
\vvec{s}_{\theta}(\vvec{y})
\equiv
-\frac{1}{\sigma}\,\vvec{\varepsilon}_{\theta}(\vvec{y}).
\label{eq:ks_score_from_noise_pred}
\end{equation}

\paragraph{Training hyperparameters (KS).}
For the results shown in the main figures we use $\sigma=0.1$ (in normalized units), Adam with learning rate $10^{-3}$, a linear warmup followed by cosine decay to $0.1\times 10^{-3}$, batch size $528$, and $100$ epochs. Each epoch is trained on a random subset of $10^5$ samples from the full trajectory to reduce compute while preserving coverage of the attractor. Batch normalization statistics are accumulated during training and frozen at inference.

\subsection{Derivation and computation of \texorpdfstring{$\bm{\Phi}$ and $\bm{\Sigma}$}{Phi and Sigma}}
\label{sec:ks_phi_sigma_derivation}

We work with the constant-matrix (mean-field) reduced Langevin model used in the main text,
\begin{equation}
\dot{\vvec{x}}(t)
=
\bm{\Phi}\,\vvec{s}(\vvec{x}(t))
+\sqrt{2}\,\bm{\Sigma}\,\vvec{\xi}(t),
\qquad
\vvec{s}(\vvec{x})\equiv \vvec{\nabla}\ln p_{\mathrm{ss}}(\vvec{x}),
\label{eq:ks_langevin_model}
\end{equation}
Define the (steady-state) correlation matrix $\bm{C}(\tau)=\langle \vvec{x}(t+\tau)\vvec{x}(t)^T\rangle$ and the Stein matrix
\begin{equation}
\bm{V}\equiv \left\langle \vvec{s}(\vvec{x})\,\vvec{x}^T\right\rangle.
\label{eq:ks_stein_matrix_def}
\end{equation}
Taking the right-derivative of $\bm{C}(\tau)$ at $\tau=0$ and using \eqref{eq:ks_langevin_model} gives
\begin{equation}
\dot{\bm{C}}(0^+)
=
\left\langle \frac{d\vvec{x}}{dt}\,\vvec{x}^T\right\rangle
=
\bm{\Phi}\,\left\langle \vvec{s}(\vvec{x})\,\vvec{x}^T\right\rangle
=
\bm{\Phi}\,\bm{V},
\label{eq:ks_Cdot_relation}
\end{equation}
where the martingale term $\sqrt{2}\bm{\Sigma}\,d\vvec{W}_t$ drops out after averaging against $\vvec{x}^T$. If $\vvec{s}$ is exact and $\vvec{x}\sim p_{\mathrm{ss}}$, then Stein's identity implies $\bm{V}=-\bm{I}$ and therefore $\bm{\Phi}=-\dot{\bm{C}}(0^+)$. In practice we do not enforce $\bm{V}=-\bm{I}$ analytically; instead we estimate $\bm{V}$ from the learned score and solve the linear system in \eqref{eq:ks_Cdot_relation}.

\subsubsection{Estimating \texorpdfstring{$\bm{V}$}{V} from the trained score network}
\label{sec:ks_V_estimator}

Because the trained network approximates the score of the \emph{perturbed} density $p_{\sigma}$, the appropriate empirical Stein matrix is
\begin{equation}
\bm{V}_{\mathrm{data}}
\equiv
\left\langle \vvec{s}_{\sigma}(\vvec{y})\,\vvec{y}^T\right\rangle_{\vvec{y}\sim p_{\sigma}}
\approx
\frac{1}{N}\sum_{n=1}^{N}\vvec{s}_{\theta}(\vvec{y}_n)\,\vvec{y}_n^T,
\qquad
\vvec{y}_n=\vvec{x}_n+\sigma\vvec{z}_n.
\label{eq:ks_V_data}
\end{equation}
For a perfectly learned score, $\bm{V}_{\mathrm{data}}\approx -\bm{I}$ provides a stringent self-consistency check; the rightmost panel of Fig.~2 in the main text reports this diagnostic.

\subsubsection{Estimating \texorpdfstring{$\dot{\bm{C}}(0^+)$}{Cdot(0)} via a finite-volume rate matrix}
\label{sec:ks_Cdot_estimator}

Direct numerical differentiation of $\bm{C}(\tau)$ at $\tau=0$ from a discrete-time trajectory is noisy and, for diffusions, sensitive to the cusp at the origin. We therefore estimate $\dot{\bm{C}}(0^+)$ from a finite-volume discretization of the Perron--Frobenius generator, following the approach described in Sec.~III. We partition the perturbed samples $\{\vvec{y}_n\}_{n=1}^{T}$ into $N_C$ control volumes $\{\Omega_j\}_{j=1}^{N_C}$ (clusters) using an adaptive tree partition with a minimum mass threshold $q_{\min}$ (in practice $q_{\min}=10^{-4}$), and encode the trajectory by the label sequence $\ell_n\in\{1,\dots,N_C\}$, where $\vvec{y}_n\in\Omega_{\ell_n}$. Let $\Delta t$ denote the sampling interval of the reduced KS time series.

From one-step transitions, we estimate a continuous-time \emph{column} generator $\bm{Q}\in\mathbb{R}^{N_C\times N_C}$ such that $\dot{\vvec{\rho}}=\bm{Q}\vvec{\rho}$ for the probability vector $\rho_j(t)\approx \mathbb{P}(\vvec{y}(t)\in\Omega_j)$. Writing $N_i$ for the number of times the trajectory occupies state $i$ (over $n=1,\dots,T-1$) and $N_{j\leftarrow i}$ for the number of observed transitions $i\to j$ over one sample (with $j\neq i$), the naive rate estimator is
\begin{equation}
Q_{j i}=\frac{N_{j\leftarrow i}}{N_i\,\Delta t}\quad (j\neq i),
\qquad
Q_{ii}=-\sum_{j\neq i}Q_{j i}.
\label{eq:ks_Q_naive}
\end{equation}
When the probability of leaving a cluster within $\Delta t$ is not small, one-step counting underestimates the true exit rates because multiple jumps can occur between observations. We correct this finite-$\Delta t$ bias by rescaling the exit-rate scale. In particular, if $p^{(1)}_{\mathrm{stay}}(i)$ denotes the empirical one-step probability to remain in state $i$, then for a continuous-time Markov chain one expects $p_{\mathrm{stay}}(i)\approx e^{Q_{ii}\Delta t}$. This yields the corrected diagonal estimate $Q_{ii}\approx \Delta t^{-1}\ln p^{(1)}_{\mathrm{stay}}(i)$ and an associated multiplicative factor
\begin{equation}
\kappa_i
\equiv
\frac{-\ln p^{(1)}_{\mathrm{stay}}(i)}{1-p^{(1)}_{\mathrm{stay}}(i)}
\qquad
\Rightarrow\qquad
Q_{j i}\leftarrow \kappa_i\,Q_{j i}\quad (j\neq i),
\label{eq:ks_finite_dt_correction}
\end{equation}
which preserves the naive destination probabilities while adjusting the overall leaving rate.
In the KS experiment we use an equivalent global ``mean-diagonal'' scaling that matches the mean corrected exit rate while keeping the sparse transition structure intact.

Let $\vvec{c}_j\in\mathbb{R}^{D}$ denote the centroid of cluster $\Omega_j$ and let $\pi_j$ denote the stationary weight (estimated empirically by $\pi_j\propto N_j$ and normalized). Define the centroid matrix $\bm{X}=[\vvec{c}_1,\dots,\vvec{c}_{N_C}]\in\mathbb{R}^{D\times N_C}$. The correlation matrix of the discretized process is
\begin{equation}
\bm{C}(\tau)\approx \bm{X}\,e^{\bm{Q}\tau}\,\mathrm{diag}(\vvec{\pi})\,\bm{X}^T,
\label{eq:ks_C_discrete}
\end{equation}
and therefore
\begin{equation}
\dot{\bm{C}}(0^+)
\approx
\bm{X}\,\bm{Q}\,\mathrm{diag}(\vvec{\pi})\,\bm{X}^T
\;\equiv\;\bm{M}.
\label{eq:ks_Cdot_discrete}
\end{equation}
This estimator is linear in $\bm{Q}$ and avoids numerical differentiation of time correlations.

\subsubsection{Solving for \texorpdfstring{$\bm{\Phi}$}{Phi} and extracting \texorpdfstring{$\bm{\Sigma}$}{Sigma}}
\label{sec:ks_phi_sigma_solver}

Combining \eqref{eq:ks_Cdot_relation} with \eqref{eq:ks_Cdot_discrete} and \eqref{eq:ks_V_data} yields the matrix equation
\begin{equation}
\bm{M}\approx \bm{\Phi}\,\bm{V}_{\mathrm{data}},
\qquad\Rightarrow\qquad
\bm{\Phi}\approx \bm{M}\,\bm{V}_{\mathrm{data}}^{-1},
\label{eq:ks_phi_solver}
\end{equation}
which we solve by a linear solve (without forming an explicit inverse). We then decompose
\begin{equation}
\bm{\Phi}_S=\tfrac12(\bm{\Phi}+\bm{\Phi}^T),
\qquad
\bm{\Phi}_A=\tfrac12(\bm{\Phi}-\bm{\Phi}^T),
\end{equation}
and identify the diffusion tensor in \eqref{eq:ks_langevin_model} with the symmetric part,
\begin{equation}
\bm{\Sigma}\bm{\Sigma}^T=\bm{\Phi}_S.
\label{eq:ks_sigma_relation}
\end{equation}
In finite data, $\bm{\Phi}_S$ may fail to be strictly positive definite; we therefore apply a minimal eigenvalue shift $\bm{\Phi}_S\leftarrow \bm{\Phi}_S+(\lvert\lambda_{\min}\rvert+\varepsilon)\bm{I}$ when needed, with $\varepsilon=5\times10^{-4}$, and take $\bm{\Sigma}$ to be the lower-triangular Cholesky factor. The matrices shown in Fig.~2 of the main text are precisely $\bm{\Phi}$, $\bm{\Phi}_S$, $\bm{\Phi}_A$, and this Cholesky factor $\bm{\Sigma}$, together with the diagnostic $\bm{V}_{\mathrm{data}}\approx -\bm{I}$.

\paragraph{Langevin integration for Fig.~1 (main text).}
Given the trained score network and the estimated $(\bm{\Phi},\bm{\Sigma})$, the reduced model trajectories are generated by Euler--Maruyama integration of \eqref{eq:ks_langevin_model}:
\begin{equation}
\vvec{x}_{n+1}=\vvec{x}_n + \Delta t_{\mathrm{EM}}\bm{\Phi}\,\vvec{s}_{\theta}(\vvec{x}_n)
 + \sqrt{2\Delta t_{\mathrm{EM}}}\,\bm{\Sigma}\,\vvec{\xi}_n,
\qquad
\vvec{\xi}_n\sim\mathcal{N}(\vvec{0},\bm{I}),
\end{equation}
with $\Delta t_{\mathrm{EM}}=5\times 10^{-3}$ and snapshots stored every $200$ steps (effective sampling interval $200\,\Delta t_{\mathrm{EM}}=1$ to match the dataset used for training and validation). Ensemble initial conditions are drawn from the empirical steady state (randomly sampled data points). The marginal PDFs and joint densities in Fig.~1 are computed from the stored samples via kernel density estimation, while ACFs are computed by averaging normalized componentwise autocorrelations over the retained modes and over ensembles.

\section{Application to Two Toy Models}
\label{appendix:toy_models}

We applied the method presented in the main text to two low-dimensional stochastic systems. These toy models are included to illustrate the method in settings where the low dimensionality allows for direct visualization and straightforward interpretation of the results. Similar systems were previously studied in Ref.~\cite{giorgini2025kgmm}. Here we demonstrate how our approach provides a stochastic model capable of reproducing both the autocorrelation functions (ACFs) and probability density functions (PDFs) directly from data. For each system, we used the estimated score function and $\bm{\Phi}$ to generate stochastic trajectories by integrating the following Langevin equation:
\begin{equation}
    \dot{\vvec{x}}(t) = \bm{\Phi} \vvec{\nabla} \ln p_{\mathrm{ss}}(\vvec{x}) + \sqrt{2}\bm{\Sigma} \vvec{\xi}(t),
\end{equation}
where $\bm{\Phi} = \bm{\Phi}_S + \bm{\Phi}_A$ is the decomposition of the drift matrix into symmetric and antisymmetric parts, $\bm{\Sigma}$ is related to $\bm{\Phi}_S$ by Cholesky decomposition, and $\vvec{\xi}(t)$ is a vector of independent delta-correlated Gaussian white noise processes.

Each system was simulated over a time interval $T \in [0,10^5 t_d]$, where $t_d$ denotes the decorrelation time of the system. These datasets were subsequently used to train the DSM-based score-function estimation method via the KGMM approach described in Sec.~\ref{appendix:score_estimation}. For each system we employed a three-layer neural network with 128 and 64 neurons in the first and second hidden layers, respectively. We used the Swish activation function between the first two layers and a linear activation function for the output layer. For each system we compared the univariate PDFs, ACFs, and trajectories obtained from the observations with those from the constructed Langevin model. When comparing trajectories, we used the same noise realizations as those used to generate the original observations, allowing for a direct pathwise comparison when the model structure permits.

\begin{figure}[htbp]
  \centering	
  \includegraphics[width=0.8\textwidth]{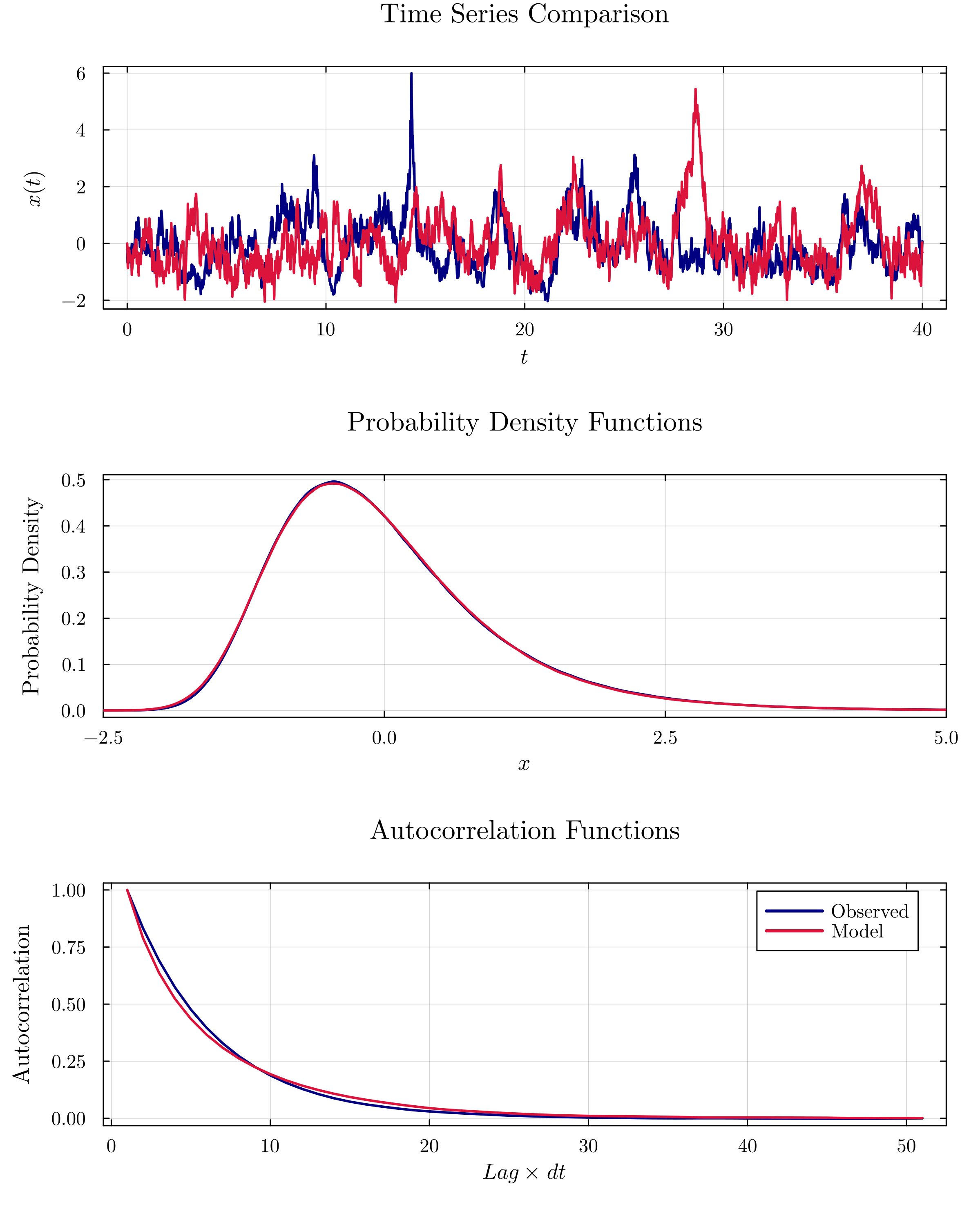}
  \caption[One-dimensional nonlinear SDE]{\textbf{One-dimensional nonlinear SDE.} \textbf{First row:} Time-series comparison between the original system (Observed) and the reconstructed dynamics (Model) using the DSM-estimated score function and $\bm{\Phi}$. The same noise realization was used to generate both time series. \textbf{Second row:} Comparison of the observed marginal PDFs (blue) with the reconstructed PDFs (red) obtained from the Langevin equation using the KGMM-estimated score function and $\bm{\Phi}$. \textbf{Third row:} Comparison of the observed ACFs (blue) with the reconstructed ACFs (red).}
  \label{fig1}
\end{figure}

\begin{figure}[htbp]
  \centering	
  \includegraphics[width=\textwidth]{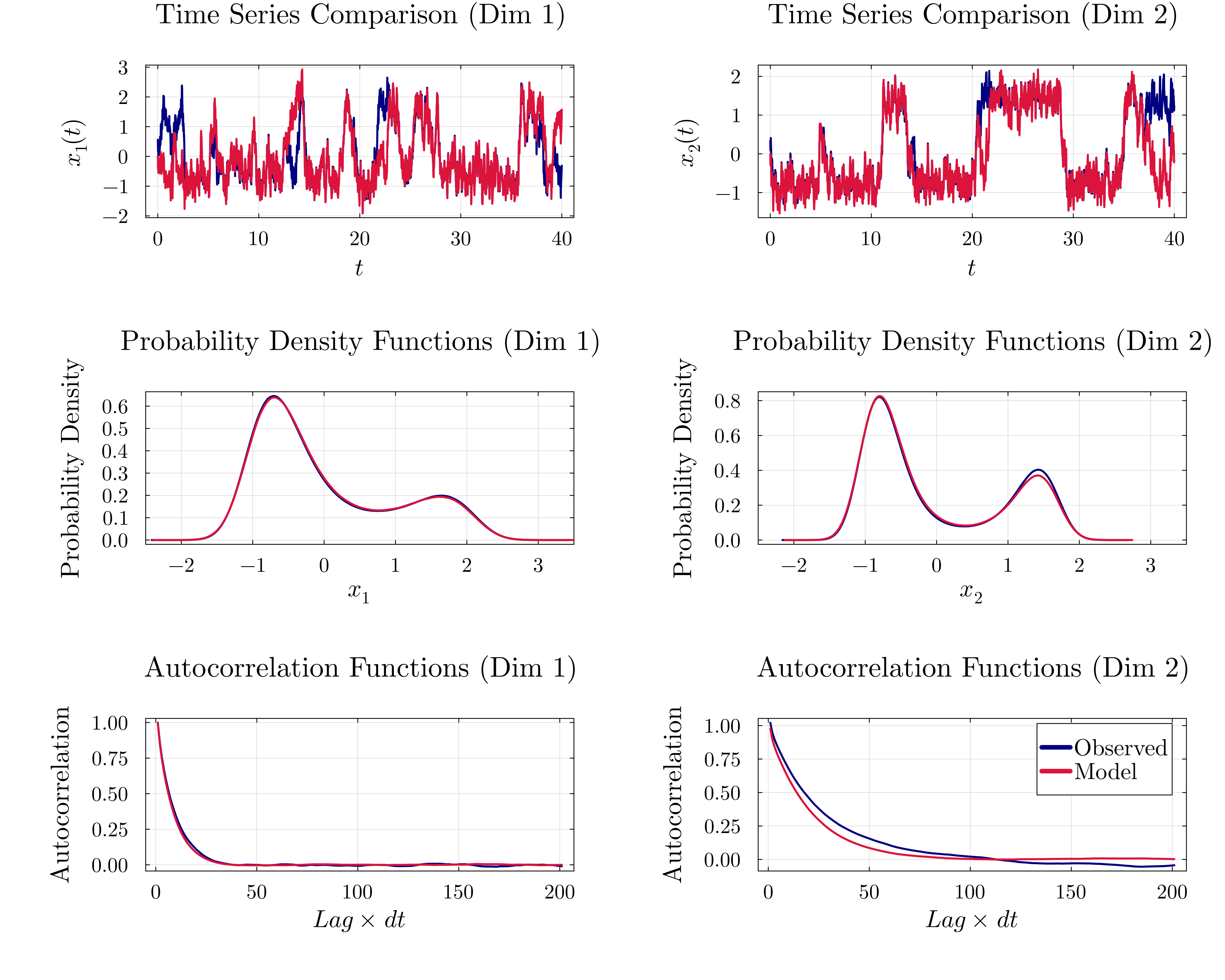}
  \caption[Two-dimensional asymmetric potential system]{\textbf{Two-dimensional asymmetric potential system.} 
	   \textbf{First row:} Comparison of trajectories between the original system (Observed) and the reconstructed dynamics (Model) using the KGMM-estimated score function and $\bm{\Phi}$. The same noise realization was used to generate both time series. \textbf{Second row:} Comparison of the observed marginal PDFs (blue) with the reconstructed PDFs (red) for each variable. 
	  \textbf{Third row:} Comparison of the observed ACFs (blue) with the reconstructed ACFs (red) for each variable.}
  \label{fig2}
\end{figure}

The two systems studied are as follows:
\begin{itemize}
    \item \textbf{One-dimensional nonlinear SDE.} This is a one-dimensional system, so the drift term has only a conservative component (antisymmetric tensors cannot exist in one dimension). The system is described by
    \begin{equation}
    \dot{x}(t) = F + a x(t) + b x^2(t) - c x^3(t) + \sigma_1\,\xi_1(t) + \sigma_2(x)\xi_2(t),
    \end{equation}
    where the coefficients are:
    \begin{equation}
    \begin{aligned}
    a &= -1.809, \quad b = -0.0667, \quad c = 0.1667, \\
    A &= 0.1265, \quad B = -0.6325, \quad F = \frac{A B}{2}, \\
    \sigma_1 &= 0.0632, \quad \sigma_2(x) = A - Bx.
    \end{aligned}
    \end{equation}
    We used $N_C=76$, $\sigma=0.05$ for the DSM algorithm.

    The method successfully reproduced both the PDF and ACF of the system, as shown in Fig.~\ref{fig1}. The reconstructed dynamics used a Langevin equation with additive noise to approximate one with multiplicative noise; consequently, despite using the same noise realization, the trajectories do not match pathwise because the noise enters the dynamics differently in the two systems.

    \item \textbf{Two-dimensional asymmetric potential system.} This system has both conservative and non-conservative components in the drift. The system is described by
    \begin{equation}
    \dot{\vvec{x}}(t) = -\bm{K}\vvec{\nabla} U(\vvec{x}) + \sqrt{2}\,\vvec{\xi}(t),
    \end{equation}
    where the potential function $U(\vvec{x})$ is
    \begin{equation}
    U(\vvec{x}) = (x_1 + A_1)^2 (x_1 - A_1)^2 + (x_2 + A_2)^2 (x_2 - A_2)^2 + B_1 x_1 + B_2 x_2,
    \end{equation}
    and the matrix $\bm{K}$ introduces a rotational component,
    \begin{equation}
    \bm{K} = \begin{pmatrix}
    1 & -0.8 \\
    0.8 & 1
    \end{pmatrix}.
    \end{equation}
    The parameters are:
    \begin{equation}
    A_1 = 1.0, \quad A_2 = 1.2, \quad B_1 = 0.6, \quad B_2 = 0.3.
    \end{equation}
    We used $N_C=761$, $\sigma=0.05$ for the DSM algorithm.
    Since fluctuations of $\bm{R}$ around its mean are zero by construction, the approximation of $\bm{R}$ with a constant antisymmetric tensor is exact. In this case the learned model coincides with the true dynamics, so we expect accurate trajectory reconstruction when the same noise realizations are used. Figure~\ref{fig2} confirms this: the trajectories obtained by integrating the reduced Langevin equation closely match the original system, and both the PDFs and ACFs are accurately reproduced.

% \item \textbf{Stochastic Lorenz-63 model.} The system is described by
%     \begin{equation}
%     \begin{aligned}
%     \dot{x}_1(t) &= \sigma (x_2(t) - x_1(t)) + \sigma_\xi\xi_1(t), \\
%     \dot{x}_2(t) &= x_1(t)(\rho - x_3(t)) - x_2(t) + \sigma_\xi\xi_2(t), \\
%     \dot{x}_3(t) &= x_1(t)x_2(t) - \beta x_3(t) + \sigma_\xi\xi_3(t),
%     \end{aligned}
%     \end{equation}
%     where the coefficients are:
%     \begin{equation}
%     \sigma = 10.0, \quad \rho = 28.0, \quad \beta = \frac{8}{3}, \quad \sigma_\xi = 5.0.
%     \end{equation}
%     We used $N_C=762$ and $\sigma=0.05$ for the DSM algorithm.
%     
%     The method performs well overall: all marginal PDFs are recovered precisely, and the ACFs for the first two coordinates match closely. For the third coordinate, the ACF does not capture the oscillatory structure but instead follows the envelope of the peaks, as shown in Fig.~\ref{fig3}. This behavior reflects the fact that the approximation of $\bm{R}(\vvec{x})$ with a constant antisymmetric matrix is not exact for this system; the constant-$\bm{R}$ closure recovers the average decay of the ACF but not its fine oscillatory features. To capture the full details of the dynamics and the complete structure of the ACF, one would need to account for the state dependence of $\bm{R}(\vvec{x})$.
\end{itemize}

% \begin{figure}[htbp]
%   \centering	
%   \includegraphics[width=\textwidth]{lorenz63.png}
%   \caption[Stochastic Lorenz-63 system]{\textbf{Stochastic Lorenz-63 system.} \textbf{First row:} Time-series comparison between the original chaotic system (Observed) and the reconstructed dynamics (Model) using the KGMM-estimated score function and $\bm{\Phi}$. The same noise realization was used to generate both time series. \textbf{Second row:} Comparison of the observed marginal PDFs (blue) with the reconstructed marginal PDFs (red) for $x_1$, $x_2$, and $x_3$. 
% 	  \textbf{Third row:} Comparison of the observed ACFs (blue) with the reconstructed ACFs (red) for each variable.}
%   \label{fig3}
% \end{figure}

\end{document}